\documentclass[a4paper,12pt,oneside]{article}
\usepackage[latin1]{inputenc}
\usepackage{amsmath,amsfonts,amssymb,bbm}
\usepackage{graphicx}
\usepackage{footnote}
\usepackage{float}
\usepackage{subfigure}
\DeclareGraphicsExtensions{.bmp,.png,.pdf,.jpg,.eps}
\usepackage{physics}
\usepackage{mathrsfs}

\usepackage[colorlinks]{hyperref}
\hypersetup{
    colorlinks=true,
    linkcolor=blue,
    filecolor=magenta,  
    urlcolor=[rgb]{0.00,0.00,0.50},    
    citecolor=red,
    }

\usepackage{url}

\advance \textheight by \topskip
\usepackage{amsmath}
\usepackage[english]{babel}
\makeatletter
 \@addtoreset{equation}{section}
\makeatother
\usepackage[latin1]{inputenc}
\usepackage{amsmath}
\numberwithin{equation}{section}

\advance \textheight by \topskip
\usepackage{amsmath}
\usepackage[english]{babel}
\DeclareMathAlphabet\mathbfcal{OMS}{cmsy}{b}{n}
\DeclareMathAlphabet{\boldmathe}{T1}{cmr}{bx}{it}
\newcommand{\mbf}[1]{\boldmathe{#1}}
\newcommand{\mbfgr}[1]{\textit{\mbox{\boldmath$#1$}}}

\def\id{\mathbbm{1}}
\def\ha{\tfrac{1}{2}}
\def\has{\tiny\frac{1}{2}}
\def\vA{\mbf{A}}
\def\vB{\mbf{B}}
\def\vK{\mbf{K}}
\def\vV{\mbf{V}}
\def\vG{\mbf{G}}
\def\vI{\mbf{I}}
\def\vJ{\mbf{J}}
\def\vL{\mbf{L}}
\def\va{\mbf{a}}
\def\vb{\mbf{b}}
\def\ve{\mbf{e}}
\def\vn{\mbf{n}}
\def\vp{\mbf{p}}
\def\vr{\mbf{r}}
\def\vs{\mbf{s}}
\def\vz{\mbf{z}}
\def\vsigma{\,\mbfgr{\sigma}}

\def\vpi{\mbfgr{\pi}}
\def\hvJ{\,\hat{\hskip-1mm\vJ}}
\def\be{\begin{equation}}
\def\ee{\end{equation}}
\def\cC{\,{\mathcal{C}}}
\def\cQ{{\mathcal{Q}}}

\def\cW{{\mathcal{W}}}

\def\Z{\mathbb Z}
\def\N{\mathbb N}
\def\R{\mathbb R}

\usepackage{todonotes}

\begin{document}
%%%%%%%%%%%%%%%%%%%%%%%%%%%%%

\title{\bf %Conformal and 
Hidden symmetry
and  (super)conformal mechanics \\
in a monopole background
%Charged particle confined to a magnetic monopole
}

\author{{\bf  Luis Inzunza, Mikhail S. Plyushchay and Andreas Wipf} 
 \\
[8pt]
{\small \textit{Departamento de F\'{\i}sica,
Universidad de Santiago de Chile, Casilla 307, Santiago,
Chile  }}\\
[8pt]
{\small \textit{Theoretisch-Physikalisches Institut, Friedrich-Schiller-Universit\"at Jena,  }}\\
{\small \textit{  Max-Wien-Platz 1,
07743 Jena, Germany}}\\
[4pt]
 \sl{\small{E-mails:   
\textcolor{blue}{luis.inzunza@usac\text{h.c}l},
\textcolor{blue}{mikhail.plyushchay@usac\text{h.c}l},
\textcolor{blue}{wipf@tpi.uni-jena.de}
}}
}
\date{}
\maketitle

\begin{abstract}
We study classical and quantum hidden symmetries of a particle with electric charge $e$ 
in the background of  a Dirac 
monopole of magnetic charge $g$ subjected 
to an additional central potential $V(r)=U(r) +(eg)^2/2mr^{2}$ 
with $U(r)=\tfrac{1}{2}m\omega^2r^2$,
similar to that 
in the one-dimensional 
conformal mechanics model of de
Alfaro, Fubini and  Furlan (AFF). 
By means of 
a non-unitary conformal bridge transformation, 
we establish a relation of  the quantum states  and of all symmetries of the system
with those of the system without harmonic trap,  $U(r)=0$.   
Introducing spin degrees of freedom via a very special spin-orbit 
 coupling,  we construct the  $\mathfrak{osp}(2\vert 2)$ superconformal extension
of the system with unbroken $\mathcal{N}=2$ Poincar\'e supersymmetry
and show  that two different superconformal 
extensions of the one-dimensional AFF model with unbroken and spontaneously broken 
supersymmetry  have a common origin.
We also show a universal  relationship
between the  dynamics of a Euclidean particle 
in an arbitrary central potential $U(r)$  and the dynamics 
of a charged particle in a monopole background 
subjected to the  potential $V(r)$.
\end{abstract}

\section{Introduction}
Hidden symmetries   
are associated with  peculiar classical and quantum properties of a system  \cite{Cariglia}. 
They are generated by higher order in  canonical momenta   integrals of motion.
 When a generator of a hidden symmetry  does  not depend  explicitly on time,  
  it transforms solutions of a system into solutions having the same energy. 
 Otherwise, a symmetry generator is the integral of motion which explicitly depends on time
and relates solutions of different energies. 
Hidden symmetries appear in a broad spectrum of the  systems,
including   the Kepler-Coulomb problem, 
anisotropic harmonic oscillator with commensurable frequencies, 
Higgs oscillator \cite{Evnin} and the Klein-Gordon equation in Anti-de Sitter space-time
\cite{Evnin+,Evnin++}, 
integrable nonlinear wave equations \cite{Integrable}, 
Calogero model \cite{CorLechPlyu}, Kerr-Newman, or more general 
Kerr-NUT-(A)dS black hole solutions of the Einstein-Maxwell equations 
\cite{FroKrtKub}. 
They also reveal  themselves nontrivially in supersymmetric extensions
of such systems, both in non-exotic  \cite{Andreas2,IKL} and exotic 
 \cite{CJNP,CJNP+}. 

One of the most known examples of 
hidden symmetries corresponds
to the case of the three-dimensional  isotropic harmonic  oscillator,
where  the closed character of the trajectories
is encoded in the Fradkin's tensor integral \cite{Frad}, 
which is  analogous to  the Laplace-Runge-Lentz vector
in the Kepler-Coulomb problem. These tensor and vector integrals  
 together with angular momentum vectors of the systems 
 define   the elliptic form of particle's trajectories and their 
 spatial orientation, and  at the quantum level 
 the presence of these integrals  
 explains  the origin of the so called ``accidental" 
 spectral degeneracy 
\cite{Pauli,Zwan}.  

Another example is provided by  reflectionless and 
finite-gap quantum systems intimately related to 
the Korteweg-de Vries and modified Korteweg-de Vries 
equations, in which the higher-derivative  Lax-Novikov integrals
separate the left- and right-moving Bloch states
and detect all the bound states and the states 
at the edges of the continuos parts (bands) 
of their spectra by annihilating them.  
Those   integrals give rise to appearance of exotic nonlinear 
supersymmetric  structures  in  super-extended versions
of such systems  \cite{CJNP,CJNP+,AraPlyu}.  

Explicitly  depending on time, dynamical 
higher order in momenta integrals of motion
appear  in rational extensions of 
one-dimensional  conformally invariant systems 
where they  detect and encode  the fine, finite-gap type  spectral structure
 \cite{CarPly2,JM,JM+,CIP,LM2}.  In these systems as well as in general case
 the higher derivative in momenta generators  of  hidden symmetries 
 give rise to non-linear generalizations of Lie algebras and 
 superalgebras \cite{BHT}.

Yet another example of the hidden symmetries corresponds 
to a non-standard extension of the fermion-monopole supersymmmetry \cite{DHoVin,DHoVin+},
the existence of which can be related to 
the Killing-Yano tensor admitted by the flat background of the monopole 
\cite{JMPH}.
In this sense its origin is similar to the origin  of the exotic ``SUSY in the sky"
of Gibbons, Rietdijk and van Holten \cite{GibRHol,Tanim,CarigliaYano},
in which additional supercharges are related to generators of hidden symmetries
as it happens  in the case of diverse black-hole solutions
of the  Einstein-Maxwell equations in (3+1) and higher dimensions \cite{FroKrtKub}.
Alternatively, exotic supersymmetry of the fermion-monopole system
finds a simple explanation in special properties  of the dynamics 
of a spin-1/2 charged particle in  monopole background \cite{Plymonosusy}. 

The flat background of the monopole is revealed 
in the dynamics of a  scalar  particle with electric charge $e$
which in its field realizes a force-free, geodesic  motion on a 
surface of a dynamical cone  
defined by the charge-monopole coupling 
parameter $\nu=eg$, where $g$ is the 
monopole's magnetic charge
  \cite{Mono1,Mono2,Mono2+}. 
One can  consider 
a  more general case of the charged particle in the monopole background 
subjected to the action of an additional 
 central potential of  the form
\begin{equation}
V(r)=\frac{\alpha}{2mr^2}+ U(r)\,,\label{pot1}
\end{equation}    
where the first term is  conformally invariant 
and $U(r)$ is a smooth function of  $r=\sqrt{\vr^2}$. 
Earlier results \cite{N4monopole} of two of us 
show that in the case  $U(r)=0$ and  particular 
value  of the  coupling $\alpha=\nu^2$,  
the projection of the particle's trajectory to  the plane ortogononal 
to the total angular momentum vector of the system 
corresponds to the one-dimensional free motion
along a straight line defined by a certain
analog  of the Laplace-Runge-Lenz vector
which for the free particle is 
$\vp\cross\vL$. It  looks  like the particle in the field of the monopole
``remembers" the integrals of motion of the system
with switched off charge-monopole coupling  ($\nu\rightarrow0$).
 Supersymmetric extension  of such a system is described 
  by the Pauli Hamiltonian of a spin-1/2 particle in
background of a self-dual or anti-self-dual dyon,
which is characterized by  a nonlinear,  quadratically extended Lie
superalgebra $D(2, 1; 1/2)$ \cite{N4monopole}
being a particular case of the exceptional 
superagebra
$D(2, 1; \alpha)$ \cite{IKL}.

From another perspective, 
in the absence of the monopole background  ($g=0$),  
two  cases of the systems described  by  potential (\ref{pot1}) 
with  $U(r)=0$ and $U(r)=\tfrac{1}{2}m\omega^2r^2$ 
are intimately  related to each other  and represent two 
forms of dynamics  in the sense  
of Dirac \cite{Dirac} corresponding to conformal symmetry.
The integrals of the system with  $U(r)=0$ 
can be obtained  by taking  some linear combinations
of integrals  of  another system and   
applying to them a limit $\omega\rightarrow 0$.
Or, in both directions the systems and their integrals  can be related 
at classical and quantum levels by a non-unitary mapping
corresponding to the conformal bridge transformation 
considered recently by us in  ref. \cite{Conformalbridge}. 
\vskip0.2cm

Based on the described relations and peculiarities, one can conjecture that 
in the presence of the monopole and  confining harmonic term $U=\tfrac{1}{2}m\omega^2r^2$ 
in potential (\ref{pot1})  something particular (related to  hidden 
and conformal symmetries) should happen in  the special case $\alpha=\nu^2$.
One could expect similar peculiar properties to be seen also in 
a superextended version of such a system. If so, it would be an interesting  
result from the point of view of  
 three-dimensional (or more generally
higher-dimensional) supersymmetric  quantum mechanics, because contrary  
to the one-dimensional case \cite{Witten,Witten+,MatSal,Cooper},
there is no canonical way to obtain such systems.  
Although there are particular and elegant constructions, see for example
\cite{Andreas2,IKL,N4monopole,Andreas1,SUSYD1,SUSYD1+,SUSYD2,SUSYD3}, it 
is in general a non-trivial task to produce  such  theories. 
In the present context, a possible  approach would be 
to look for a general $(3+0)$ dimensional Dirac type operator
as a supercharge (square root) of a Klein-Gordon type ``super-Hamiltonian".

\vskip0.1cm
This  work is devoted to the investigation of the conjectures   specified  
in the previous paragraph, and 
in conclusion of this section we describe the organization of the paper 
and briefly summarize  its results.

\vskip0.1cm

In Section \ref{ClassicalSection}
we  investigate the classical theory of a
charged scalar particle in a monopole background 
subjected to the action of additional scalar potential 
of the form (\ref {pot1}) with a harmonic trap 
$U(r)=\tfrac{1}{2}m\omega^2 r^2$, which has a 
 nature similar  to 
the  potential in one-dimensional
conformal mechanics model of de Alfaro, Fubini and Furlan (AFF) 
\cite{AFF}
described by the Hamiltonian\footnote{This model and its supersymmetric 
extensions \cite{SCM2,SCM5} play important role, in particular,  
in  black hole physics \cite{BlackHold1,BlackHold2,BlackHold3,BlackHold4}, 
AdS/CFT correspondence \cite{AdSCFT1,AdSCFT2}, 
cosmology \cite{Cosmo1,Cosmo2}, 
and holographic QCD \cite{App1}.}
 \begin{equation}
\label{AFFmodel}
H_{\text{AFF}}=\frac{p^2}{2m}+\frac{m\omega^2q^2}{2}+\frac{\ell(\ell+1)}{2m q^2}\,.
\end{equation}
We solve the equations of motion and 
find that the trajectories are closed for an arbitrary choice of 
initial conditions only in the special case 
when $\alpha=\nu^2$. We show that in this special case the 
dynamics of the radial variable is governed  by conformal
Newton-Hooke symmetry of the system and 
compare it with the dynamics of the system with 
$U(r)=0$ studied earlier in details in ref. \cite{N4monopole}.
It turns out that in  the special case $\alpha=\nu^2$
the full dynamics -- including its angular part --
is controlled by a hidden symmetry described by 
the integrals of motion of order four in momenta variables.
These integrals  define the orientation of the trajectory projected to 
the plane orthogonal to the conserved total angular momentum 
vector. In spite of the fourth order in momenta nature of 
generators of the hidden symmetry, in contrast with 
 the second order generators for
the isotropic harmonic oscillator, they 
reveal a structure somehow similar to  the Fradkin's tensor
in the latter system.
Section \ref{QuantumSection} is devoted to
the quantum theory  of the system with 
$\alpha=\nu^2$, where, in particular, with the 
help of the hidden symmetry 
we identify  the full set of its ladder operators.
Using the results of 
our previous paper \cite{Conformalbridge}, 
we   also construct the conformal bridge transformation 
which relates the quantum spectrum of our system 
with that of the model 
without confining harmonic term as has been studied in 
\cite{N4monopole}. Particularly, we show that 
coherent states for the present system 
are generated by the conformal bridge transformation 
 from  non-normalizable energy eigenstates of the 
system with $U(r)=0$, while all its energy eigenstates 
are produced from Jordan states of zero energy 
of the latter system.
This transformation also establishes a relation 
between symmetries of both systems, including 
generators of their hidden symmetries.
 In Section \ref{SUSYsection}
we consider a supersymmetric  generalization  of 
the quantum system by introducing a very special 
spin-orbit coupling, that  allows us to obtain 
the  $\mathfrak{osp}(2\vert 2)$ superconformal extension
of the system with unbroken $\mathcal{N}=2$ Poincar\'e supersymmetry.
We also demonstrate that two different superconformal 
extensions of the one-dimensional AFF model with unbroken and spontaneously broken 
phases of $\mathcal{N}=2$ Poincar\'e supersymmetry have a common origin 
in the three-dimensional $\mathfrak{osp}(2\vert 2)$ superconformal symmetry 
of the spin-1/2 particle  in a monopole background.
When switching off the monopole background 
by setting $g=0$,  the non-relativistic limit of the Dirac oscillator  
considered in refs. \cite{Balen,DiracOs,RevLett,QueMosh,Quesne} is recovered, and 
the  $\mathfrak{osp}(2\vert  2)$ superconformal symmetry  remains
intact. On the other hand, 
when switching off the harmonic trap 
by taking $\omega=0$, the superconformal Hamiltonian
of our extended system takes the form of
the  Pauli type Hamiltonian for a charged spin-1/2 particle in a field of the
self-dual dyon studied in \cite{N4monopole}. 
The discussion of our results and  an outlook 
are presented in Section  \ref{DiscussionSection}, where
we also generalize the observation of Section~\ref{ClassicalSection} 
by showing a universal  relationship
between the three-dimensional dynamics of a Euclidean particle 
in an arbitrary central potential $U(r)$  and the dynamics 
of a charged particle in a monopole background 
subjected to the action of the central potential $U(r)+\nu^2/2mr^2$.
 Several technical details are  moved to four appendices.

\section{
Conformal mechanics  in a monopole background
}
\label{ClassicalSection}
In this section we study 
the dynamics of a  
charged particle 
in background of a magnetic monopole 
in the presence of an  additional central potential which is 
a three-dimensional analog of  that in the AFF
conformal mechanics  model \cite{AFF}. 
The system we investigate is given  by the Hamiltonian 
\begin{equation}
\label{ClassicalH}
H=\frac{\vpi^2}{2m}+\frac{m\omega^2 r^2}{2}+\frac{\alpha}{2mr^2}\,, 
\end{equation}
where $\omega>0$, $\vpi=\vp-e\vA$,   $\vA$  
is a U(1) gauge potential  of a
Dirac magnetic monopole  at the origin 
with charge $g$, 
$\nabla\cross\vA=\vB=g\vr/r^3$, and the coupling
$\alpha$ should be chosen appropriately to prevent a fall to the center, see below.
We solve the Hamiltonian equations, 
study the  conformal Newton-Hooke symmetry
of the system, and investigate a hidden symmetry
which appears in  a special case   
 $\alpha=\nu^2$, $\nu=eg$. 
 We follow here the line of reasoning used in \cite{N4monopole} to 
 identify the hidden symmetry
 and characterize the  particle's  trajectories.

\subsection{Classical dynamics}

The particle's coordinates and  kinetic momenta obey the 
Poisson brackets  relations
\begin{equation}
\{r_i,\pi_j\}=\delta_{ij}\,,\qquad\{r_i,r_j\}=0\,,\qquad \{\pi_i,\pi_j\}=e\epsilon_{ijk}B_{k}\,,
\end{equation}
which give rise to the equations of motion
\begin{equation}
\label{CanonEq}
\dot{\vr}=\frac{1}{m}\vpi\,,\qquad
\dot{\vpi}=\frac{1}{mr^3}(\alpha\vn-\nu\,\vr
\times\vpi)-m\omega^2\vr\,,
\end{equation}
where $\vn={\vr}/{r}$.
From  (\ref{CanonEq}) we  derive the  equations  
\begin{equation}
\label{eqmotion2}
\frac{d r}{dt}=\frac{1}{m}\pi_r\,, \qquad
\dot{\vn}=\frac{1}{mr^2}\,\vJ\cross\vn\,,
\end{equation}
where we denote $\pi_r=\vn\cdot\vpi$,  and 
\begin{equation}
\label{ClassicPoincare}
\vJ=\vr\cross\vpi-\nu\vn
\,
\end{equation}
is the conserved Poincar\'e vector  
identified as the 
angular momentum  of the system,
\begin{equation}
\{J_i,J_j\}=\epsilon_{ijk}J_k\,,\qquad
\{J_i,r_j\}=\epsilon_{ijk}r_k\,,\qquad
\{J_i,\pi_j\}=\epsilon_{ijk}\pi_k\,. 
\end{equation}
From (\ref{ClassicPoincare}) it folllows that $\vJ\cdot \vn=-\nu$ and
$\vJ^2\geq \nu^2$,
i.e. a trajectory of the particle lies on
the  surface of a cone with symmetry axis given by
the angular momentum vector $\vJ$ and cone's angle 
\begin{equation}\label{thetacone}
\theta=\arccos(-\nu/J)\,, \qquad J=\sqrt{\vJ^2} \,.
\end{equation}
In the limit case $J^2=\nu^2$ 
the cone degenerates into a half-line. 
If $\nu<0$, then $\theta=0$ and  the particle moves  on  a
half-line directed along the angular momentum
$\vJ$, whereas  $\theta=\pi$ if  $\nu>0$ and
the particle moves  on a half-line opposite to the direction 
of the vector $\vJ$.

By means of Eq.  
(\ref{ClassicPoincare}),  the Hamiltonian  can be presented
in the form
\begin{equation}\label{HAFF}
H=\frac{\pi_r^2}{2m}+\frac{\mathscr{L}^2}{2mr^2}+\frac{m\omega^2r^2}{2}\,,
\qquad\mathscr{L}^2
:=\vJ^2-\nu^2+\alpha\,,
\end{equation}
which 
shows that the radial dynamical variables $r$ and $\pi_r$,  $\{r,\pi_r\}=1$, 
behave like $q$ and $p$ in the one dimensional AFF model (\ref{AFFmodel}). 
From (\ref{HAFF}) it follows that there is no fall to the center
if $\mathscr{L}^2> 0$, i.e. $\alpha>0$,
 that we will assume from now on. 
Eq.   (\ref{HAFF}) also implies that 
 the possible values of the angular  momentum  $J$ and energy 
 obey the relation
\begin{equation}
\label{in_HL}
\frac{\mathscr{L}\omega}{H}
:=
\lambda \leq 1\,.
\end{equation}

\begin{figure}[H]
\begin{center}
\includegraphics[scale=0.2]{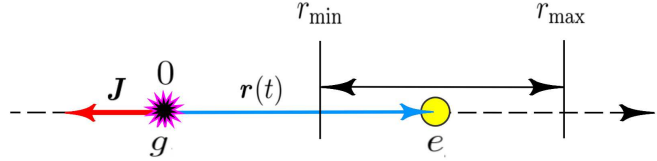}
\caption{\small{Illustration for the case $J^2=\nu^2$ with   $\nu=eg>0$. 
The monopole with charge $g$ is at the origin of the coordinate system, 
and the vectors  $\vJ$ and $\vr$ are oriented in opposite directions.
The position of the particle of charge $e$ oscillates 
 between $r_\mathrm{min}$ and $r_\mathrm{max}$. For $\nu<0$  the vector $\vJ$ 
 is oriented in the same direction as $\vr$.  } }
\label{figure0}
\end{center} 
\end{figure}

Let $r_0=r(t_0)$  corresponds to a turning point,
 $\dot r(t_0)=0$. Then according to  relation $\pi_r=m\dot{r}$ and    (\ref{HAFF}), 
 $r_0^2$ is defined by the  
 equation
\begin{equation}
r_0^4-\frac{2H}{m\omega^2}r_0^2+\frac{\mathscr{L}^2}{m^2\omega^2}=0\,.
\end{equation} 
Its  two solutions 
\begin{equation}
\label{returning1}
r_{\pm}^2=\frac{H}{m\omega^2}(1\pm \rho)\,,\qquad 
0\leq \rho=\sqrt{1- \lambda^2}<1\,,
\end{equation}
satisfy the relation
\begin{equation}
\label{r+r-}
r_+r_-=\frac{\mathscr{L}}{m\omega}\,.
\end{equation}
If,  
for simplicity,
we choose  the initial moment of time $t_0=0$ such that 
$r(0)=r_-=r_\mathrm{min}$, 
integration of the first equation in (\ref{eqmotion2}) with taking into account 
Eq.  (\ref{HAFF}) yields  
\begin{equation}
\label{r(t)1}
r^2(t)=\frac{H}{m\omega^2}(1- \rho\cos(2\omega t))\,.
\end{equation}   
So, $r(t)$ oscillates   between $r_\mathrm{min}=r_-$
and $r_\mathrm{max}=r_+$ with a period $\pi/\omega$. 
The particular case with $H=\mathscr{L}\omega$ corresponds to 
a circular motion for which $r(t)=r_+=r_-=(\mathscr{L}/m\omega)^{1/2}$.
If $J^2=\nu^2$, the particle realizes one-dimensional oscillations 
 (\ref{r(t)1})  between $r_\mathrm{min}$ and $r_\mathrm{max}$ lying on the half-line 
 specified below Eq. (\ref{thetacone}),
see Fig. \ref{figure0}

To solve the vector equation in (\ref{eqmotion2})
in the case $J^2>\nu^2$, which we will assume in what follows, 
we decompose  $\vn$ into  the component  parallel to the angular momentum
and the orthogonal component,
\begin{equation}
\label{n(phi)}
\vn(t)=
\vn_\parallel+\vn_{\bot}(t)
=-\nu\,\frac{\hvJ}{J}
+\vn_{\bot}(t),\qquad \vJ\cdot\vn_{\bot}(t)=0\,,
\end{equation}   
where $\hvJ$ is the unit vector in 
the
direction of $\vJ$.
Since the parallel component $\vn_\parallel$ is constant, we 
conclude that the
orthogonal component $\vn_{\bot}(t)$  has constant length 
and
thus 
describes
a circle in the plane orthogonal to $\vJ$, 
\begin{equation}\label{nparaort}
\vn_{\bot}(t)=\vn_{\bot}(0)\cos\varphi(t)
+\hvJ\cross\vn_{\bot}(0)\sin\varphi(t)\,.
\end{equation}
Using the second equation in (\ref{eqmotion2}) we then obtain
\begin{equation}\label{dotphi}
\dot{\varphi}=\frac{J}{mr^2}\,,
\end{equation}
that
yields the time-dependence of the angular variable\footnote{
Eqs. (\ref{nparaort}) and (\ref{dotphi}) imply a rotation of $\vn_{\bot}$
in the positive, clockwise
direction looking on it from the direction of the vector $\vJ$.
If $\vJ$ is oriented along  $\ve_z$, and $\nu<0$, $0<\theta< \pi/2$ in (\ref{thetacone}),
the particle's trajectory lies on the upper sheet of the cone
and $\vn_{\bot}$ rotates in a clockwise direction in the horizontal plane.
If $\vJ$ is oriented along  $-\ve_z$, and $\nu>0$,  $\pi/2<\theta< \pi$,
then the trajectory lies again on the upper sheet of the cone, but the 
 vector $\vn_{\bot}$ rotates   anti-clockwise in the $(x,y)$ plane if to look
 at it from $\ve_z$.}.
Integrating this equation with using  (\ref{r(t)1})
and assuming $\varphi(0)=0$,
we 
obtain
\begin{eqnarray}
\label{phi(t)} 
&\varphi(t)=\frac{J}{\mathscr{L}} \arctan(\frac{r_\mathrm{max}}{r_\mathrm{min}}\tan(\omega t))\,.&
\end{eqnarray}
It is 
convenient to parametrize the orbits
by expressing $\xi=1/r^2$ as a
function of $\varphi$. We have
$d\xi/d\varphi=-2r^{-3}\dot r/\dot{\varphi}$, 
and so, 
\begin{equation}
\label{Angular eq}
\frac{d\xi}{d\varphi}=-\frac{2}{J}
\sqrt{2mH \xi-m^2\omega^2-\mathscr{L}^2\xi^2}\,.
\end{equation}
Integration of this equation yields 
\begin{eqnarray}
\xi(\varphi)=\frac{1}{r^2(\varphi)}=\frac{mH}{\mathscr{L}^2}\left[1+\rho\cos(\frac{2\mathscr{L}}{J}\varphi)\right]\,,
\end{eqnarray}
which 
corresponds to 
$r(\varphi=0)=r_\mathrm{\min}$ 
and  
the angular 
period $\pi J/\mathscr{L}$. 
The condition for a periodic trajectory is
\begin{equation}
\frac{2\mathscr{L}}{J}2\pi l_r=2\pi l_a\quad
\Longleftrightarrow
\quad
\frac{2\mathscr{L}}{J}=\frac{l_a}{l_r},
\qquad l_r,l_a=1,2,\ldots\,.
\end{equation}
From the definition of $\mathscr{L}$  in (\ref{HAFF}) 
we find  that the trajectories are closed for arbitrary 
values of $J$ if  and only if  
$\alpha=\nu^2$. 
If $\alpha\neq \nu^2$, 
the trajectory will be closed only for 
special values of the angular momentum given by
the condition
\begin{equation}
\label{alpha nu J}
\alpha=\nu^2
+\left(\frac{1}{4}\frac{l_a^2}{l_r^2}-1\right)J^2\,,
\end{equation}
 and in this case Eq.  (\ref{in_HL}) takes the form 
$\frac{l_a}{l_r}\leq \frac{2H}{\omega J}$.

\begin{figure}[H]
\begin{center}
\includegraphics[scale=0.350]{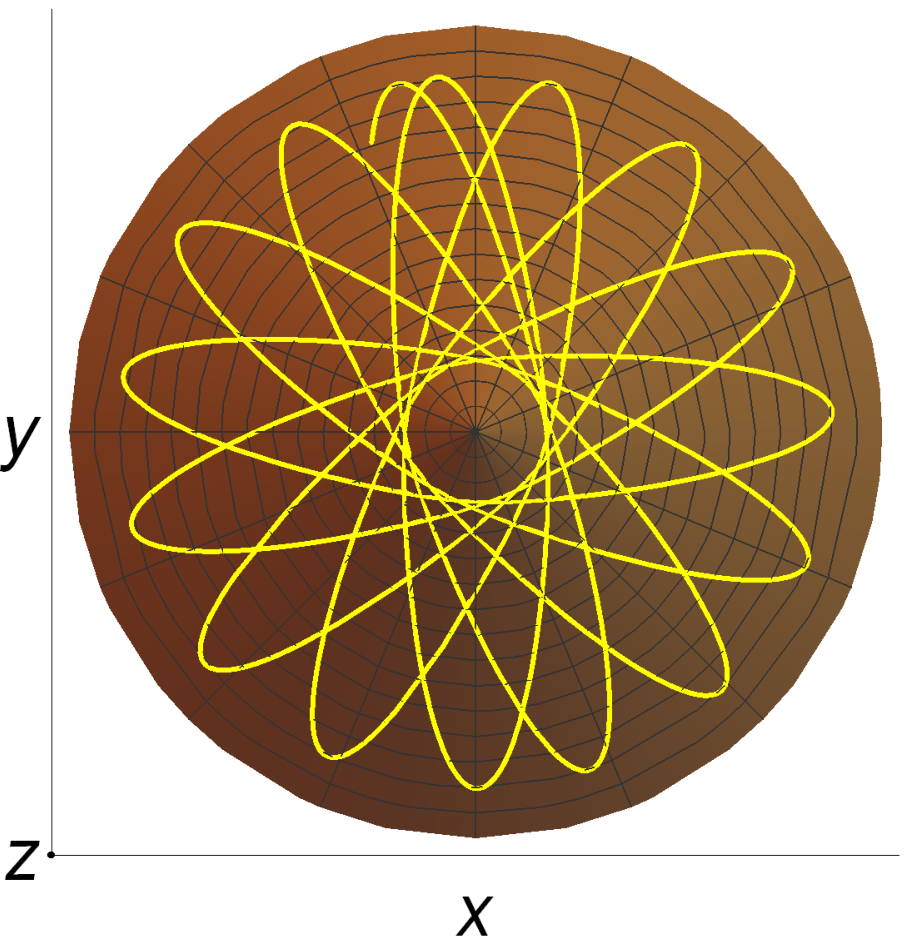}\hskip8mm
\includegraphics[scale=0.350]{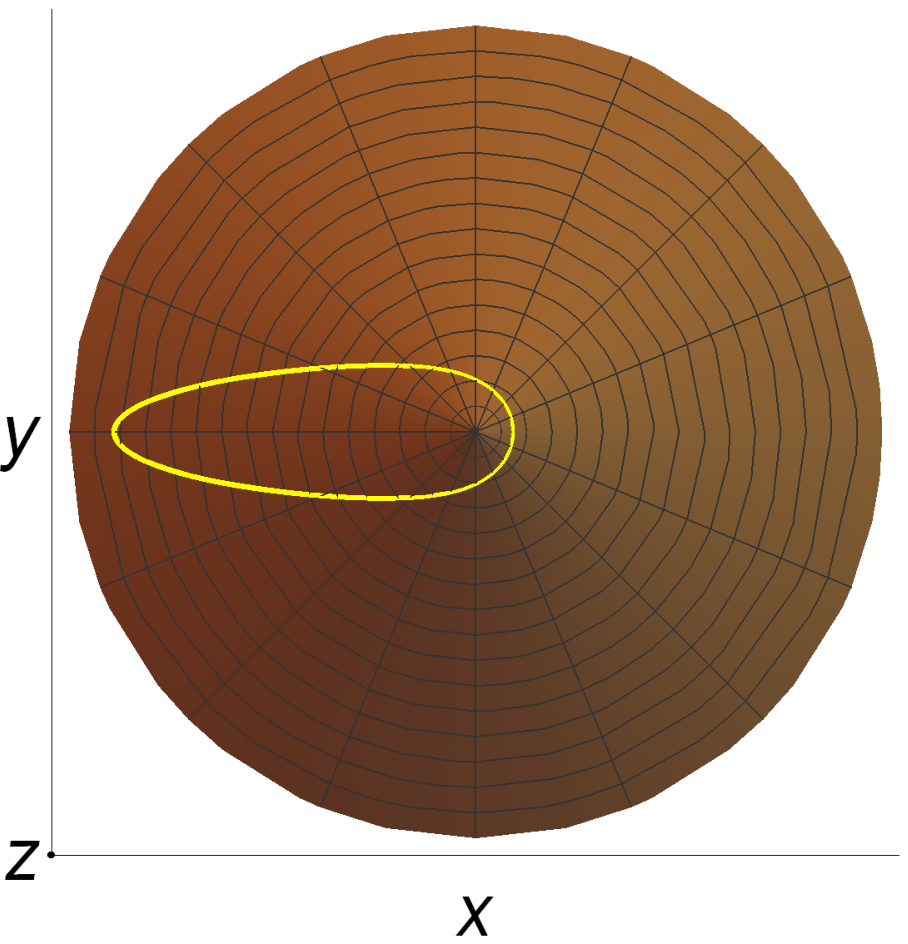}\hskip8mm
\includegraphics[scale=0.350]{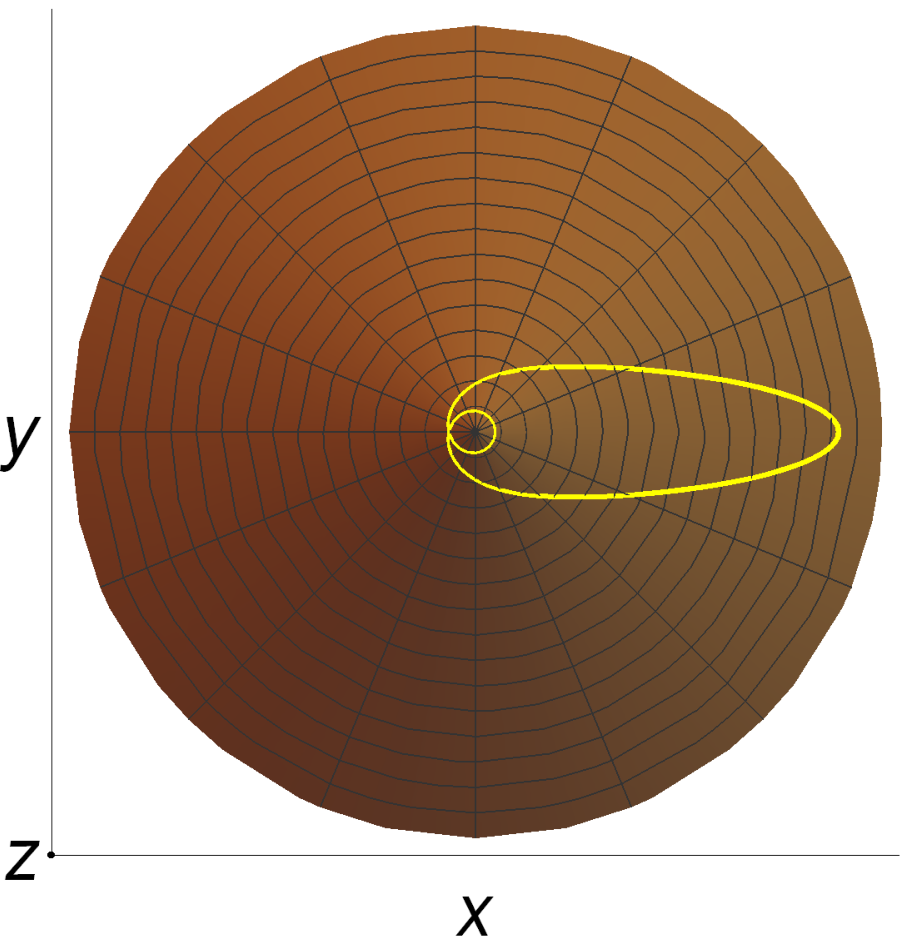}\hskip10mm
\includegraphics[scale=0.350]{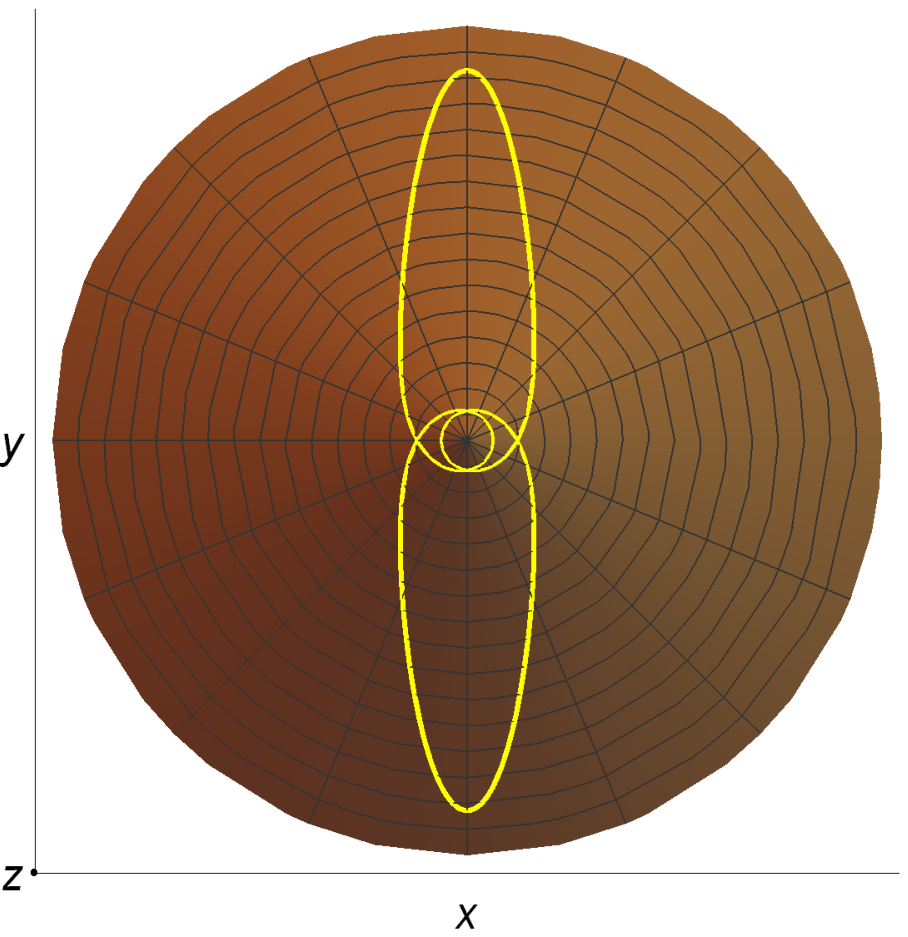}\\\vskip8mm
\includegraphics[scale=0.350]{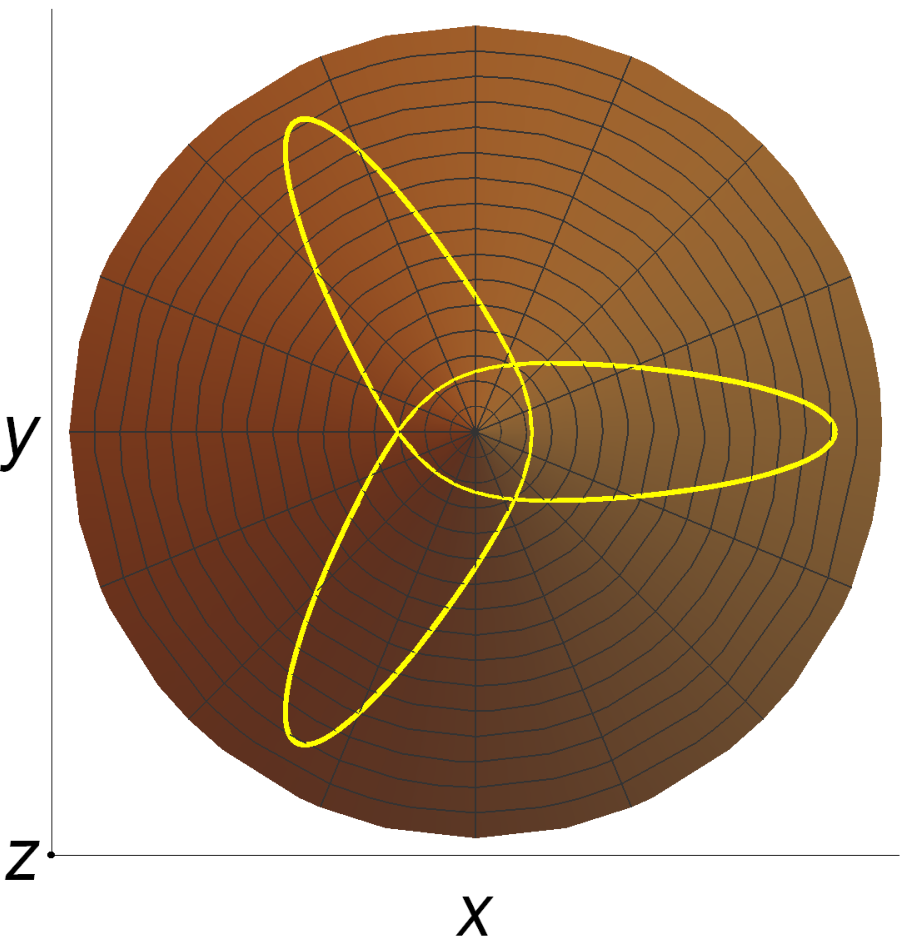}\hskip8mm
\includegraphics[scale=0.350]{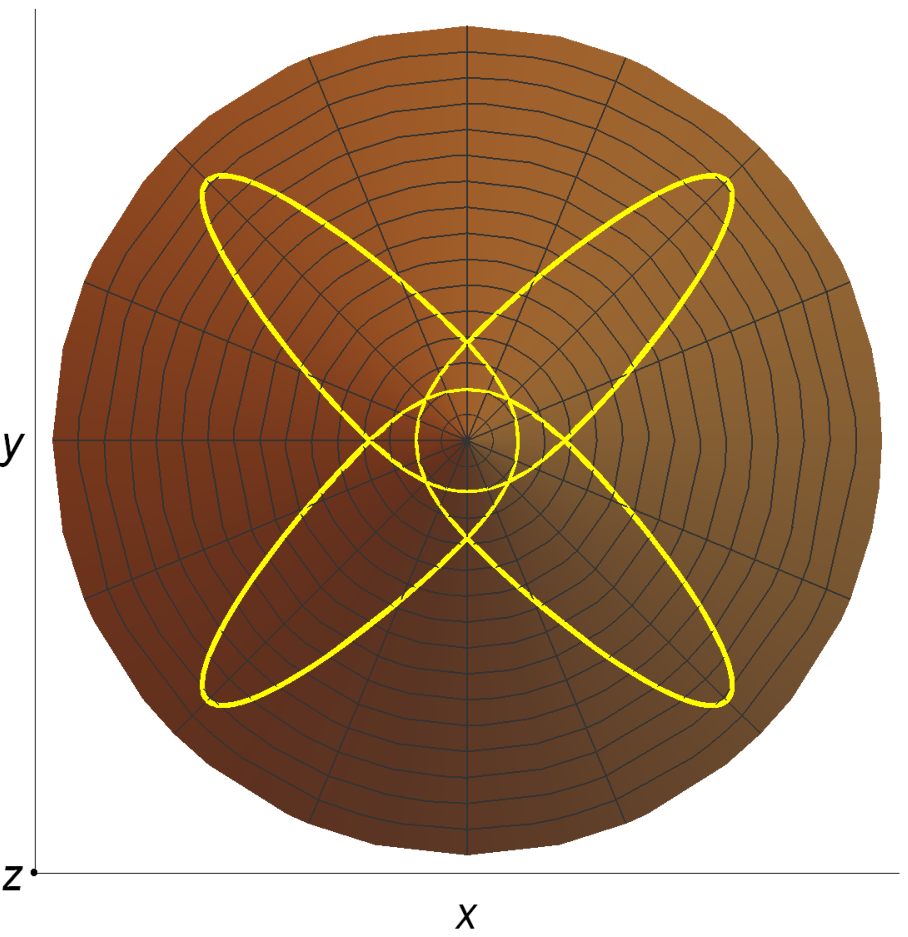}\hskip8mm
\includegraphics[scale=0.350]{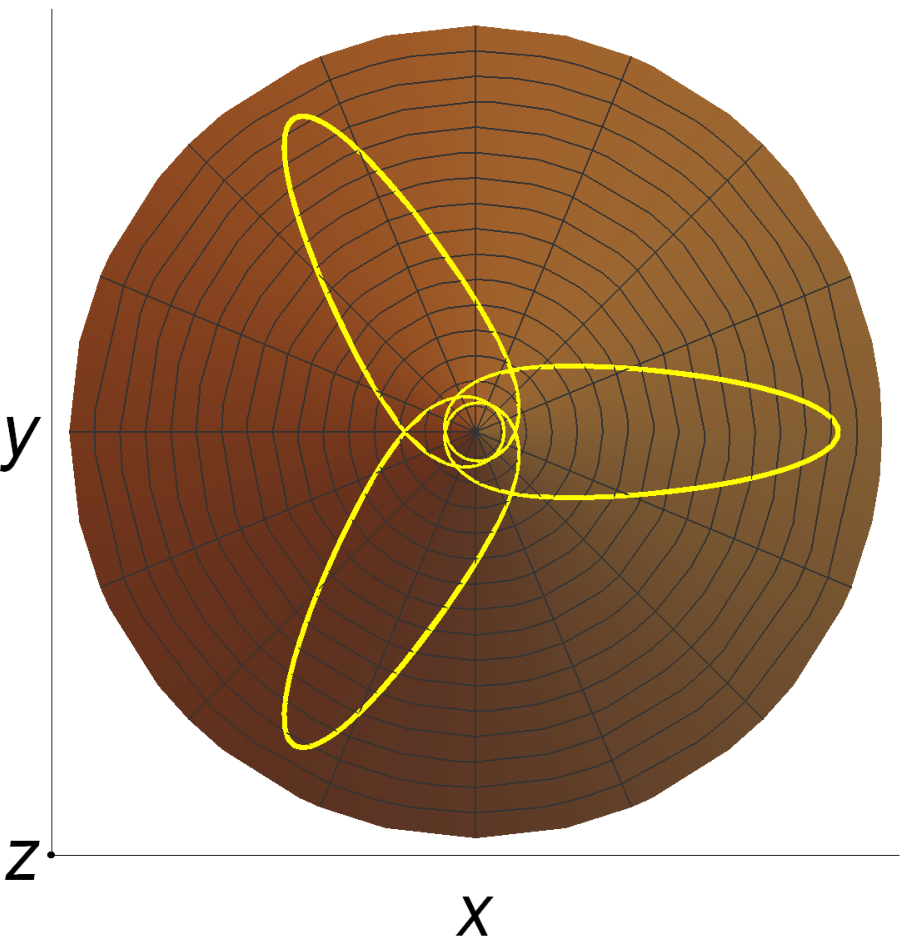}\hskip8mm
\includegraphics[scale=0.350]{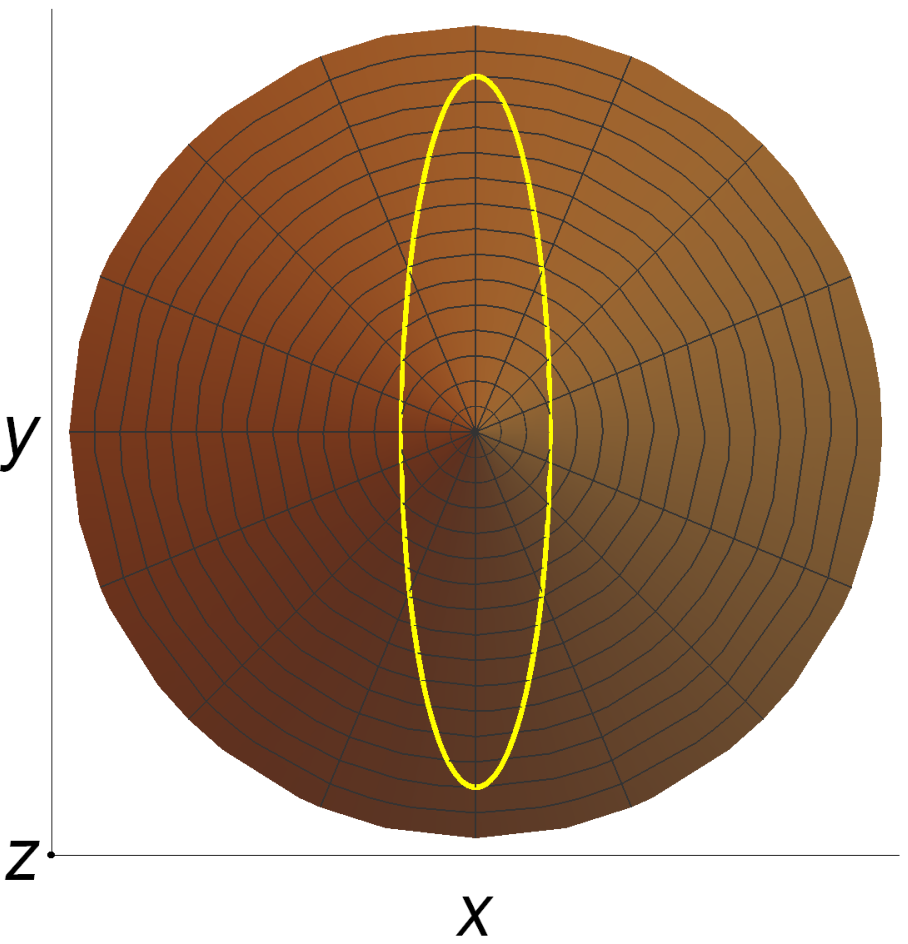}
\caption{\small{
The depicted trajectories  correspond to the vector 
$\vJ$ oriented along $\ve_z$.
The first figure in the top row represents the generic case with
non-closed trajectory. 
The other figures are examples of closed trajectories
with parameters satisfying  the relation  (\ref{alpha nu J}),
with quotients  $l_a/l_r=\{1,\,1/2,\,2/3,\,3/2,\,4/3,\,3/4,\,2\}$ are sequentially shown.  
The last}
relation  $l_a/l_r=2$ corresponds to  the special case $\alpha=\nu^2$. }
\label{figure1}
\end{center} 
\end{figure}

Figure \ref{figure1} illustrates several particular orbits
lying on the corresponding conical surface 
in a general case $\alpha\neq \nu^2$ and 
in the special case $\alpha= \nu^2$.
Trajectories  $\vr(\varphi)$ are shown there 
for fixed values of $H$, $\vJ$ and $\nu$, but for different 
 values of $\alpha$. 
Below we shall see that when  $\alpha=\nu^2$,
the projection to the plane  orthogonal to $\vJ$ 
 of the trajectory shown on the last 
plot  is an ellipse centered at 
the origin of the coordinate system 
similarly to the case of the three-dimensional 
isotropic harmonic oscillator.
This corresponds to a fundamental universal 
property   of the magnetic monopole background which
we discuss in the last section.
Since the center of the projected elliptical trajectory is
in the center of an ellipse, the angular period $P_a$ is twice 
the radial period $P_r$, $P_a/P_r=2$,  similarly to the
isotropic harmonic oscillator. This is different from the 
picture of the finite orbits in Kepler problem
where the force center is in one of the foci, and as a result
$P_a=P_r$. This similarity with the isotropic oscillator 
and contrast to the Kepler problem are also reflected
in the spectra of the systems at the quantum level.

\subsection{Conformal Newton-Hooke symmetry}
In this subsection we derive the conformal Newton-Hooke 
symmetry \cite{NH1,NH2,NH3,NH4} for the  system  (\ref{ClassicalH})
and compare it with the conformal symmetry of the model
with $\omega=0$ studied in \cite{N4monopole}. 

Using the AFF form of the Hamiltonian (\ref{HAFF}),
one can show  that the  complex quantity
 \begin{equation}
 \label{Cladder}
\cC=e^{2i\omega t}\left(\frac{\pi_r^2}{2m}+\frac{\mathscr{L}^2}{2mr^2}-\frac{m\omega^2r^2}{2}- 
i\omega r\pi_r\right)=
 e^{2i\omega t}\left(H-m\omega^2r^2 - i\omega r\pi_r\right)\,,
\end{equation}   
and its complex conjugate $\cC^*$
are explicitly depending on time  integrals of motion
which together with $H$ generate the 
$\mathfrak{sl}(2,\R)$ algebra
\begin{equation}
\{H,\cC\}=2i\omega \cC\,,\quad 
\{H,\cC^*\}=-2i\omega \cC^*
\,,\quad 
\{\cC,\cC^*\}=-4i\omega H\,.
\end{equation} 
In terms of $\cC$ and $\cC^*$, the generators of the   
Newton-Hooke symmetry are given by
\begin{equation}
\label{KyD}
D=\frac{i}{4\omega}
(\cC-\cC^*)\,,\qquad
K=\frac{1	}{4\omega^2}
(2H-\cC-\cC^*)\,,
\end{equation}
and together with $H$ they satisfy the algebra 
\begin{equation}
\label{NH}
\{D,H\}=H-2\omega^2K\,,\qquad
\{D,K\}=-K\,,\qquad
\{K,H\}=2D\,,
\end{equation}
whose  Casimir invariant is  
\begin{equation}\label{clasCas}
\mathscr{F}= D^2+\omega^2K^2-KH=-\frac{1}{4}\mathscr{L}^2\,.
\end{equation}
In terms of these generators, the function   $r^2(t)$  is presented in the form
\begin{equation}\label{rDKH}
r^2(t)=\frac{2}{m\omega^2}\big(\omega D\sin(2\omega t)+\omega^2K\cos(2\omega t)+H\sin^2(\omega t)\big)\,.
\end{equation}
The values of the dynamical integrals $D$ and $K$ depend on the choice of initial conditions 
for $\vr$ and $\dot{\vr}$,
and as follows from (\ref{Cladder}) and (\ref{KyD}), our choice $r(0)=r_\mathrm{min}$, $\dot{r}(0)=0$ corresponds to
$D=0$ and $K=\frac{1}{2}mr_\mathrm{min}^2$. 
For these values of $D$ and $K$, (\ref{rDKH}) takes  the form 
(\ref{r(t)1}). On the other hand, since $\vJ\cdot\vr(t)=-\nu r(t)$,
for a general choice of initial conditions  Eq. (\ref{rDKH}) 
shows  that the dynamics  of the projection of $\vr(t)$
on the direction of the conserved angular momentum is controlled 
by the conformal  Newton-Hooke  symmetry of the system.
According to (\ref{rDKH}), the oscillation period
of $r(t)$ is $\pi/\omega$, and taking into account 
the value of the Casimir invariant, one can 
check that in the general case Eq. (\ref{rDKH}) also implies  that $r(t)$  oscillates 
between the values $r_\mathrm{min}$ and $r_\mathrm{max}$
given by Eq. (\ref{returning1}).

\begin{figure}[H]
\begin{center}
\includegraphics[scale=0.350]{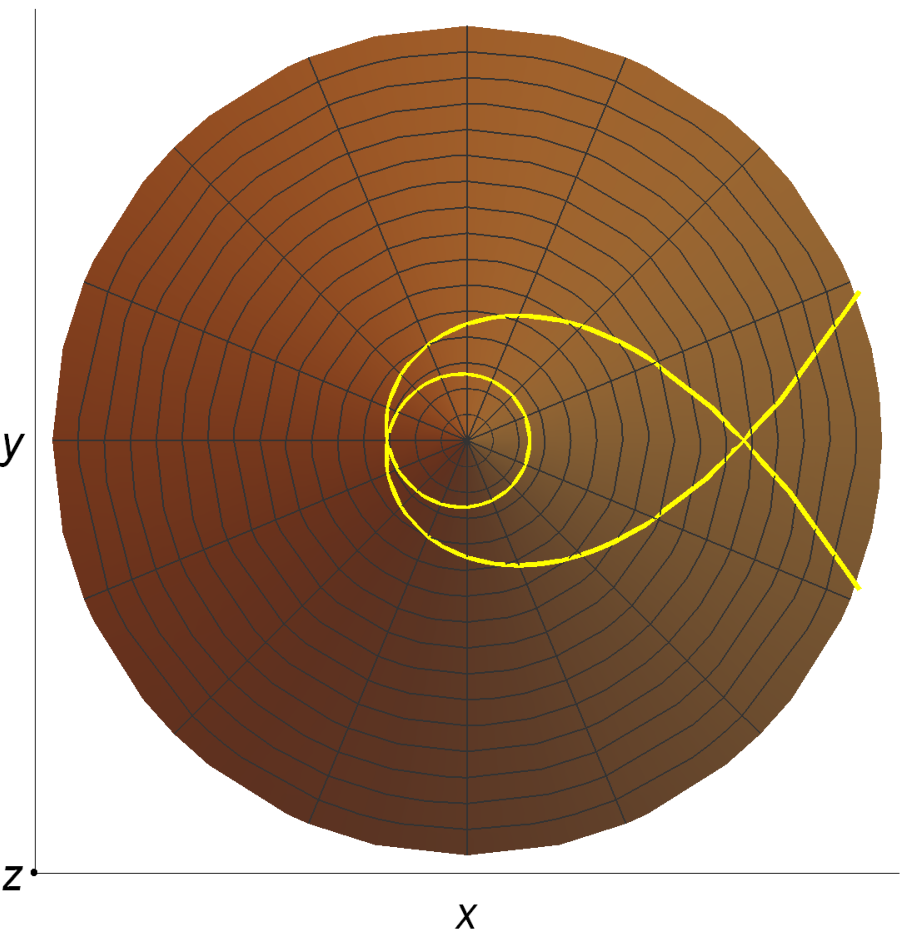}\hskip8mm
\includegraphics[scale=0.350]{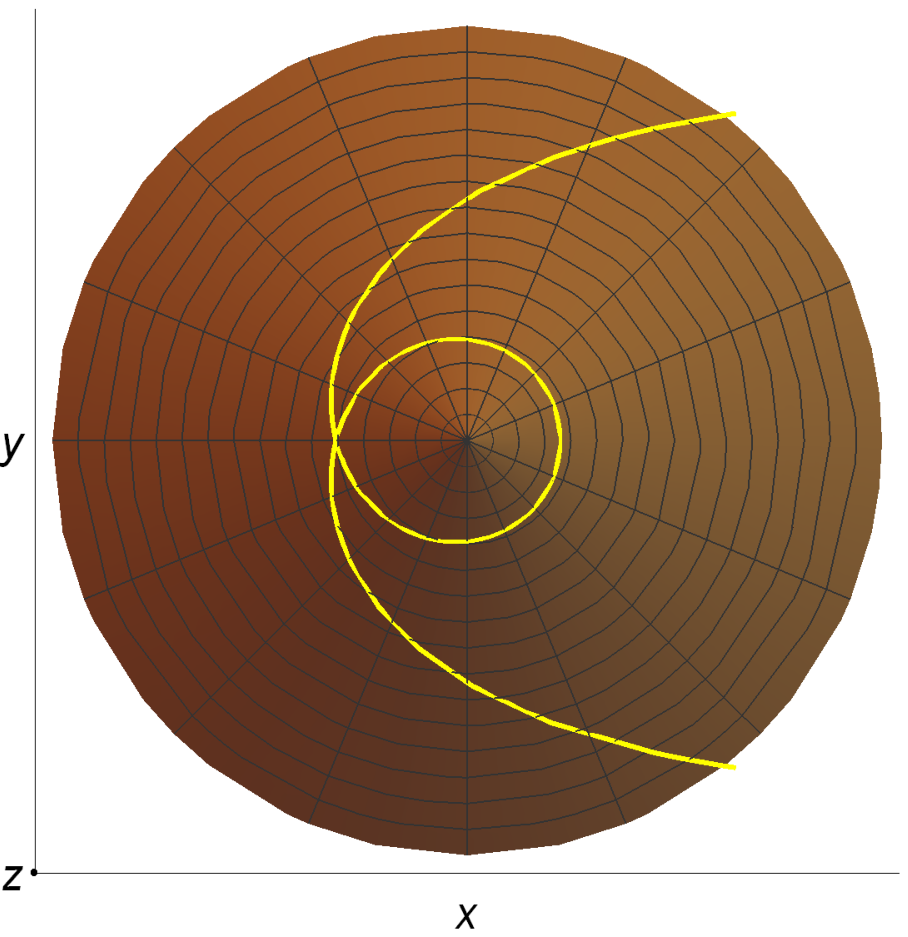}\hskip8mm
\includegraphics[scale=0.350]{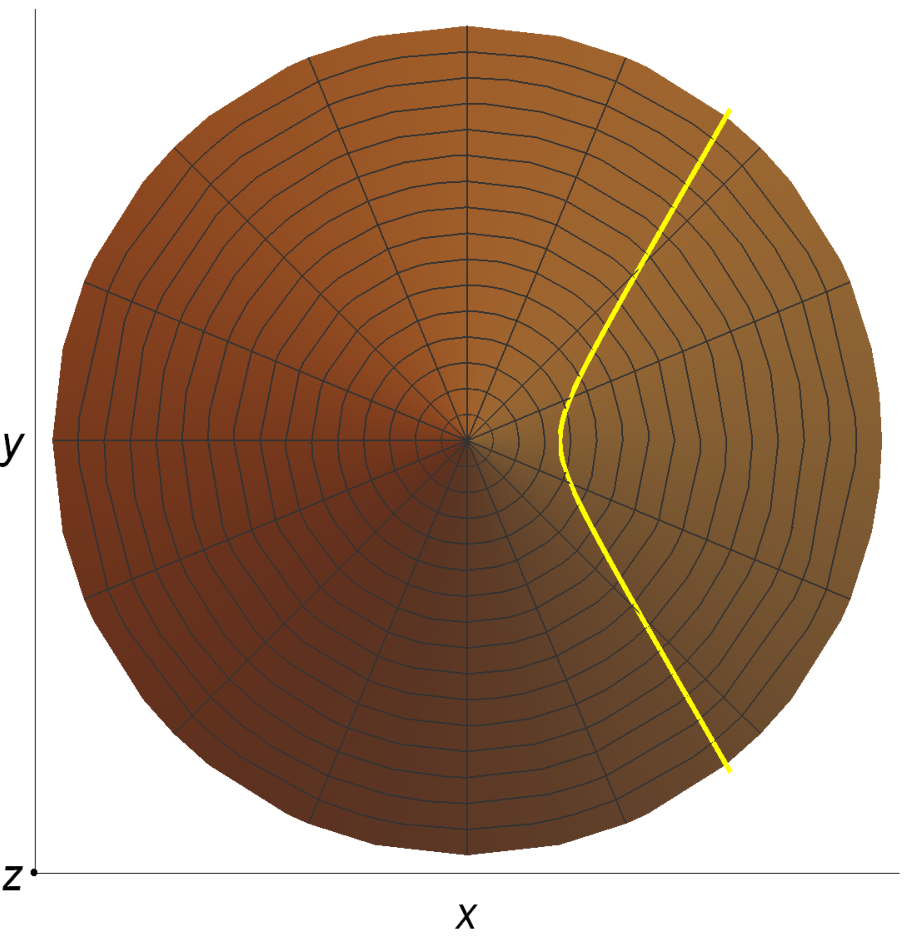}\hskip10mm
\includegraphics[scale=0.350]{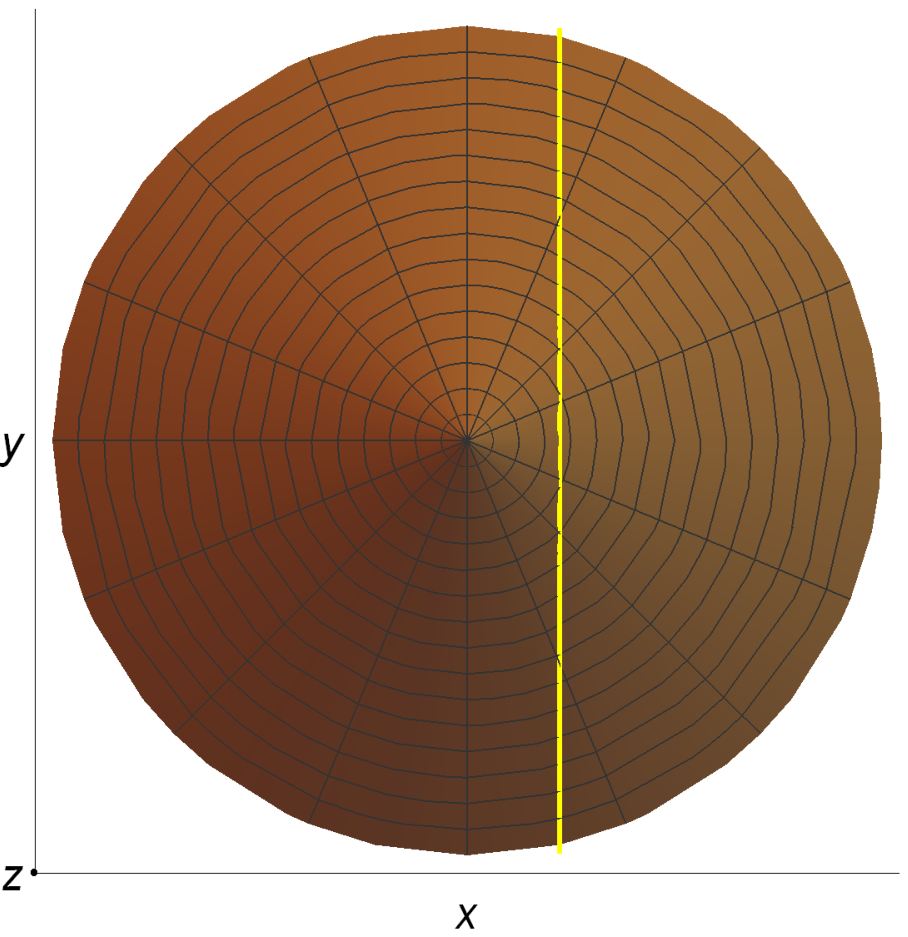}
\caption{\small{Each plot represents a trajectory for a specific  value of 
$\alpha$ chosen according to (\ref{alpha nu J}) 
with the vector $\vJ$ oriented along  $\ve_z$. 
From left to right the cases $l_a/l_r=\{3/2,\,1/2,\,3,\,2\}$ are shown, 
where the last plot 
corresponds to the special case  $\alpha=\nu^2$.} }
\label{figure3}
\end{center} 
\end{figure}

To conclude this part of the  analysis,
we comment on  the limit 
$\omega\rightarrow 0$. In this case the generators $H$, $D$ and $K$
take the form 
\begin{equation}
\label{freemotion}
H_0=\frac{\pi_r^2}{2m}+\frac{\mathscr{L}^2}{2mr^2}\,,\qquad
D_0=\frac{1}{2}r\pi_r-H_0t\,,\qquad
K_0=\frac{mr^2}{2}-Dt-H_0t^2\,,
\end{equation}
and  satisfy the conformal algebra 
\begin{equation}
\{D_0,H_0\}=H_0\,,\qquad
\{D_0,K_0\}=-K_0\,,\qquad
\{K_0,H_0\}=2D_0\,. 
\end{equation}
The case $\alpha=0$ of the system $H_0$ 
corresponds  to a geodesic motion on the dynamical cone \cite{Mono2,Mono2+}.
The special case of $\alpha=\nu^2$, on the other hand,  was studied in
\cite{N4monopole}. It was shown there that 
the trajectory of the particle, projected 
 to the plane orthogonal to
$\vJ$, is  a straight line along which the projected particle's motion 
takes place with constant velocity. Consistently with these
peculiar properties,  in the special case 
$\alpha=\nu^2$ the system with $H_0$ 
possesses a hidden symmetry  described by 
the integral of motion $\vV=\vpi\times\vJ$ being a sort of 
Laplace-Runge-Lenz vector, in the plane orthogonal to which 
and parallel to $\vJ$ 
the particle's trajectory lies  \cite{N4monopole}.
In Fig. \ref{figure3} some plots of the trajectories are shown 
for the 
system (\ref{freemotion}). 

\subsection{
The case $\alpha=\nu^2$\,: hidden symmetry}\label{hidclassym}

In the case $\alpha=\nu^2$ the particle described
by the Hamiltonian (\ref{ClassicalH}) admits
additional  integrals  of motion 
responsible for the closed nature of the trajectories 
for arbitrary choice of initial conditions. 
The integrals  are derived by an algebraic
approach  as in Fradkin's construction 
for the isotropic  three-dimensional harmonic oscillator 
\cite{Frad}.

Let us first consider arbitrary values of $\alpha$
assuming, as we indicated above, $J^2>\nu^2$.
Our first step is to introduce the vector quantities  
\begin{align}
\vI_1&=\vpi\cross\vJ\cos(\omega t)+\omega m \vr\cross\vJ\sin(\omega t)\,,
\label{I1}\\
\vI_2&=\vpi\cross\vJ\sin(\omega t)-\omega m \vr\cross\vJ\cos(\omega t)\,.
\label{I2}
\end{align}
The time-dependence of these vectors in the plane orthogonal 
to $\vJ$ follows from the time-dependence of $\vpi$ and $\vr$
in (\ref{CanonEq}),
\begin{equation}
\label{canonicaleqI1I2}
\frac{d \vI_1}{dt}=\frac{1}{mr^3}(\alpha-\nu^2)\vn\cross\vJ\cos(\omega t)\,,\qquad
\frac{d \vI_2}{dt}=\frac{1}{mr^3}(\alpha-\nu^2)\vn\cross\vJ\sin(\omega t)\,.
\end{equation}
Then
\begin{equation}
\label{scalarproduct}
\vI_1\cdot \vI_2 =\frac{1}{2}(\nu^2-\alpha)(J^2-\nu^2)
\frac{\sin(2\omega t)}{2r^2}\,,
\end{equation}
where we have taken into account our choice of the initial condition
$\pi_r(0)=0$.
The initial values of these vectors are
\begin{equation}\label{I(0)}
\vI_1(0)=\frac{J^2}{r_\mathrm{min}}
\vn_{\bot}(0)\,,\qquad
\vI_2(0)=m\omega r_\mathrm{min}
\vJ\cross\vn_{\bot}(0)\,.
\end{equation}
Thus,  $\vI_1$ and $\vI_2$   are orthogonal to each other at $t=0$,
but in general case of $\alpha\neq \nu^2$ 
their scalar product is not zero and changes 
periodically with period $\pi/\omega$.

For the particular choice $\alpha=\nu^2$, the vectors  
$\vI_1$ and $\vI_2$ are orthogonal vector integrals of motion of order 2 
in the kinetic momenta, and  so, they correspond to 
the ``hidden symmetries" 
\cite{Cariglia} of the system.
They are, however,  dynamical, explicitly time-dependent
integrals of motion
(similarly  to  generators of the 
conformal Newton-Hook symmetry $D$ and $K$),
$\frac{d}{dt}\vI_{1,2}=\frac{\partial \vI_{1,2}}{\partial t}
+\{\vI_{1,2},H\}=0$.
Their lengths are also dynamical integrals 
whose values, again  in the sense  of a total time derivative,
take constant values 
\begin{equation}
\vert\vI_{1}\vert
=m\omega\sqrt{J^2-\nu^2}\,r_\mathrm{max},\qquad
\vert\vI_{2}\vert
=m\omega\sqrt{J^2-\nu^2}\,r_\mathrm{min}\,,\label{lengthint}
\end{equation}
where we have taken into account Eqs. (\ref{I(0)}) and (\ref{r+r-}).
The sum of their squares, however, is a true integral of motion 
whose value is a function of $H$ and $J$,
\be\label{I12+I22}
\vI_1^2+\vI_2^2=2mH(J^2-\nu^2)\,.
\ee
These vectors point
in the direction of the semi-axes of the elliptic trajectory
in the plane orthogonal to $\vJ$.  
The  lengths of semi-major and semi-minor axes 
correspond to  those of the vectors $r\vn_{\bot}(0)$ and 
$r\hvJ\cross \vn_{\bot}(0)$,  and are equal to 
$r_\mathrm{max}\sqrt{1-\nu^2/J^2},$ and  
$r_\mathrm{min}\sqrt{1-\nu^2/J^2}$.
We note that
in  general case
$\alpha\neq \nu^2$  the periodic change  of the scalar
product of $\vI_1$ and $\vI_2$ 
implies a precession of the orbit, see 
Fig. \ref{figure1}. 

Let us now investigate in more detail the most interesting case $\alpha=\nu^2$ given by 
the Hamiltonian 
\begin{equation}
\label{ClassicHnu}
H=\frac{\vpi^2}{2m}+\frac{m\omega^2}{2}r^2+\frac{\nu^2}{2mr^2}\,.
\end{equation}
To express the general solution in terms of the conserved $\vJ$ and 
dynamical integrals $\vI_1$ and $\vI_2$ in (\ref{I1})
and (\ref{I2}), which for $\alpha=\nu^2$ become true integrals of motion,
we note that
\begin{eqnarray}
\vr(t)\cross\vJ= \frac{1}{m\omega}(\vI_{1}\sin(\omega t)-\vI_{2}\cos(\omega t) )\,,\qquad
\vpi(t)\cross\vJ=\vI_{1}\cos(\omega t)+\vI_{2}\sin(\omega t)\,.\label{r(t)}
\end{eqnarray}
By means of  the relations 
\begin{equation}
\label{relationJyr}
\vJ\cross(\vr(t)\cross\vJ)=J^2\vr(t)+\nu r(t)\vJ\,,\qquad 
|\vr\cross\vJ|^2=(J^2-\nu^2)r^2(t)\,,\qquad  
\end{equation}
we can express the position $\vr(t)$ of the particle
as follows,  
\begin{equation}
\label{r(t)2}
\vr(t)=\frac{1}{m\omega J^2}\left( \vJ\cross \vI_{1}\sin \omega t-\vJ\times\vI_{2}\cos \omega t
-\nu \frac{\sqrt{I_1^2\sin^2\omega t+I_2^2\cos^2\omega t}}{\sqrt{J^2-\nu^2}}\vJ\right)\,,
\end{equation} 
with $\vI_{1}=\vI_1(0)$ and  
$\vI_{2}=\vI_2(0)$. This yields us 
$\vr(t)$
 and kinetic momentum $\vpi=m\dot{\vr}$
at any given time presented in terms of the angular 
momentum   and dynamical vector  integrals.
\vskip0.1cm

Alternatively, one can follow a more algebraic approach to 
extract information on the trajectories without explicitly 
solving the equations of motion.
It is well known from the seminal paper of Fradkin \cite{Frad}
that for the three-dimensional isotropic harmonic oscillator all 
symmetries of the trajectories are encoded in a tensor integral 
of motion. In the remainder of this 
subsection
we construct 
an analogous tensor for the system at hand to find 
the
trajectories by a
linear algebra techniques.
We begin with the  tensor integrals
\begin{equation}
T^{ij}=T^{(ij)}+T^{[ij]}\,,\qquad
T^{(ij)}=\frac{1}{2}(I_{1}^{i}I_1^j+I_{2}^i I_2^j)\,,\qquad
T^{[ij]}=\frac{1}{2}(I_{1}^{i}I_2^j-I_{1}^j I_2^i)\,.\label{symm_tensor}
\end{equation}
They,  unlike the vectors $\vI_1$ and $\vI_2$,  but like
the quadratic expression (\ref{I12+I22}) are the true,
not depending explicitly on time integrals of motion,
$\frac{d}{dt}T^{ij}=\{T^{ij},H\}=0$,
whose explicit form in phase space variables is
\be
2T^{ij}=(\vpi \times \vJ)^i (\vpi \times \vJ)^j+
m^2\omega^2 (\vr \times \vJ)^i (\vr \times \vJ)^j
+
\epsilon^{ijk}m\omega(J^2-\nu^2)J_{k}\,.
\ee
In accordance with (\ref{I12+I22}),
their  components satisfy relations
\begin{equation}
\text{tr}(T)= m(J^2-\nu^2)H\,,\qquad
\epsilon_{ijk}T^{[jk]}=m\omega(J^2-\nu^2)J_{i}\,.
\end{equation}
As the anti-symmetric part of $T^{ij}$ is related with the Poincar\'e integral, 
we only need to use the symmetric part $T^{(ij)}$, which 
is related but not identical to Fradkin's tensor.
Since the vectors (\ref{I1}), (\ref{I2}) are orthogonal to each other and
to $\vJ$,  we immediately conclude that
$\vJ,\vI_1$ and $\vI_2$ are eigenvectors of $T^{(ij)}$ with eigenvalues equal,
respectively,
 to zero and 
\begin{align}
\lambda_1=\vert\vI_1\vert^2&=\frac{1}{2}m^2\omega^2(J^2-\nu^2)r^2_\mathrm{max}\,,\\
\lambda_2=\vert\vI_2\vert^2&=\frac{1}{2}m^2\omega^2(J^2-\nu^2)r^2_\mathrm{min}\,,
\label{eigenvalues}
\end{align}
where we have taken into account (\ref{lengthint}).
The relations 
\begin{equation}
\vI_1\cdot \vr=(J^2-\nu^2)\cos(\omega t)\,,\qquad
\vI_2\cdot \vr=(J^2-\nu^2)\sin(\omega t)\,,
\end{equation}
allow us to conclude 
that the quadratic form $\vr^T T\vr$
is time-independent,
\begin{equation}
2r_i T^{ij}r_j=(\vI_1\cdot\vr)^2+(\vI_2\cdot\vr)^2=(J^2-\nu^2)^2\,.
\label{quad_form}
\end{equation}
In a coordinate system with orthonormal base $\ve_x=\hat{\vI}_1,\ve_y=\hat{\vI}_2$ and
$\ve_z=\hat{\vJ}$, the quadratic form (\ref{quad_form})
simplifies to
\begin{equation}
\lambda_1 x^2+\lambda_2 y^2=(J^2-\nu^2)^2\,.
\end{equation}
With $r_\mathrm{max}r_\mathrm{min}=J/(m\omega)$
one ends up with the equation for an ellipse
in the plane orthogonal to $\vJ$:
\begin{equation}
\frac{x^2}{r^2_\mathrm{min}}+\frac{y^2}{r^2_\mathrm{max}}=\frac{J^2-\nu^2}
{J^2}
\,.
\end{equation}
The lengths of the semi-major axis  and semi-minor axis  of
the ellipse are 
$r_\mathrm{max}\sqrt{1-\nu^2/J^2}$,
$r_\mathrm{min}\sqrt{1-\nu^2/J^2}$ 
in accordance with that was found above.

For 
 quantum theory it is of advantage to use the
complex form of dynamical integrals of motion
\begin{equation}
\label{ladders1}
\va=\frac{1}{\sqrt{2}}
(\vI_1+i\vI_2)=\vb\cross\vJ\, e^{i\omega t}\,,
\quad
\vb=\frac{1}{\sqrt{2}}
(\vpi-i\omega m \vr)\,,
\end{equation}
and its complex conjugate $\va^*$.
They satisfy the non-linear Poisson bracket relations  
\begin{eqnarray}
&
\{H,\va\}=i\omega \va\,,\qquad
\{J_i,a_j^\#\}=\epsilon_{ijk}a_k^\#\,,\qquad
\{a_i^\#,a_j^\#\}=-m\,\cC^\#\epsilon_{ijk}J_k\,,
&\label{PossonBrak-a1}\\&
\{a_{i}^*,a_{j}\}=im\omega[(2J^2-\nu^2)\delta_{ij}-J_{i}J_j)]-mH\epsilon_{ijk}J_k\,,
& \label{PossonBrak-a2}
\end{eqnarray}
and
are related to the generators 
(\ref{Cladder})
of
the conformal symmetry, 
\begin{eqnarray}
a_{i}^\# a_{i}^\#=m(J^2-\nu^2)\cC^\#\,,
\end{eqnarray}
where $a^\#_i$ denotes either $a_i$ or $a^*_i$, and similarly for $\cC$, 
$\cC^*$. 
The $a^\#_i$ and $\cC^\#$ are classical analogues of 
the ladder operators in the related quantum system.
In terms of $a^\#_i$ the tensor integrals $T_{(ij)}=T^{(ij)}$ 
and
$T_{[ij]}=T^{[ij]}$ take the form 
\begin{equation}
\label{Tequa}
T_{(ij)}=\tfrac{1}{2}\big(a_i^* a_j+a_j^* a_i\big),\qquad 
T_{[ij]}=\tfrac{i}{2}\big(a_i a_j^*-a_j a_i^*\big)=\tfrac{1}{2}m\omega(J^2-\nu^2)\epsilon_{ijk}J_k\,.
\end{equation}
In fact, such kind of tensors 
were 
considered in earlier
studies of the quantized system by Vinet  \emph{et al }\cite{Vinet}. We find
it useful to exploit these integrals for the classical system
at hand.

Symmetric tensor integral $T_{(ij)}$ satisfies 
the Poisson bracket reations
\be
\{J_i,T_{(jk)}\}=\epsilon_{ijl}T_{(lk)}+\epsilon_{ikl}T_{(jl)}\,,
\ee
and 
\begin{equation}
\{T_{(ij)},T_{(lk)}\}=m(\epsilon_{ils}\mathcal{F}_{jk}+\epsilon_{iks}\mathcal{F}_{jl}+
\epsilon_{jls}\mathcal{F}_{ik}+\epsilon_{jks}\mathcal{F}_{im})J_s\,,
\end{equation}
where 
\be
\mathcal{F}_{ij}=\tfrac{1}{4}m\omega^2(J^2-\nu^2)^2\delta_{ij}-HT_{(ij)}\,,
\ee
and we have used  Eqs.  (\ref{PossonBrak-a1}), (\ref{PossonBrak-a2}) 
and the equality 
\begin{equation}
\cC^*a_ia_j+a_i^*a_j^*\cC=2HT_{(ij)}+m\omega^2(J^2-\nu^2)[J^2\delta_{ij}-J_iJ_j]\,.
\end{equation}

In the following section we shall see that the
quantum analog of the classical integrals of motion
$\cC^\#$ and $\va^\#$, due to their dynamical  nature  of conservation,
provide us with 
a complete set of the spectrum generating operators for the quantum system with 
Hamiltonian (\ref{ClassicHnu}).

To conclude
this section, we comment on the limit $\nu\rightarrow0$,
when we recover the isotropic harmonic oscillator.
In this limit the integral $\va$ and its complex conjugate 
reduce
to the vector product of the orbital angular 
momentum and the classical analogs of the first order ladder 
operators. Instead of considering
the dynamical integral $\va$ 
one may choose the vector  
\begin{equation}
\vz=\big(\vb+\frac{\nu}{J^2}(\vb\cdot\vn)\vJ\big)e^{i\omega t}=\frac{1}{J^2}\vJ\cross\va\,
\end{equation}  
with $\vb$ defined in  (\ref{ladders1}).
This integral  and its complex conjugate, 
which indeed contain the same physical information as $\va$ and $\va^\#$, fulfill the Poisson bracket relations    
\begin{equation}
\{\vz_i^\#,\vz_k^\#\}=-\frac{\nu^2}{2J^4}\cC^\#\epsilon_{ikl}J_l\,,\qquad
\{\vz_i,\vz_k^*\}=-im\omega\delta_{i,k}+\frac{1}{J^4}\mathcal{O}(\nu^2)\,,
\end{equation}   
where $\mathcal{O}(\nu^2)$ are terms of order $\nu^2$. In the limit $\nu\rightarrow0$ they are just the classical analogs of the 
first order ladder operators satisfying
the Heisenberg algebra. However, the appearance of 
the non-local operator $1/J^2$ in quantum mechanics 
complicates the analysis considerably and we 
prefer to use the
integrals  (\ref{ladders1}) to deal with local
operators  in what follows.

\section{Quantum case for $\alpha=\nu^2$}
\label{QuantumSection}
The quantum theory of the system with Hamiltonian
(\ref{ClassicHnu}) has been studied earlier 
in \cite{Mcin,Vinet}. Here we reconsider the system
as a preparation for our investigation 
of the related superconformal system with spin-orbit coupling
in the next section, 
and to discuss an interesting relation 
of the generalized quantum AFF system in the monopole
background with its analog without a confining harmonic potential term.
First we solve the Schr\"odinger equation by
a separation of variables and afterwards we solve the problem of ladder operators 
by exploiting 
the quantum conformal symmetry as well as the hidden symmetry. 
In a separate subsection,
we connect this result with the quantum version of the system 
$H_0$ in (\ref{freemotion}) by 
construction of 
a  
``conformal bridge transformation"   
 following ref. \cite{Conformalbridge}. 
We shall use the units in which $m=1$ and $\hbar=1$.
\vskip0.1cm

In coordinate representation the basic 
commutation relations  are 
\begin{equation}
[\hat{r}_i,\hat{r}_j]=0\,,\qquad [\hat{r}_i,\hat{\pi}_j]=i\delta_{ij}\,,\qquad
[\hat{\pi}_i,\hat{\pi}_j]=i\nu\epsilon_{ijk}\frac{\hat{r}_k}{r^3}\,.
\end{equation}
In what follows we shall skip the hat symbol $\,\hat{{}}\,\,\,$
to simplify the notation.  
The Hamiltonian 
(\ref{ClassicHnu}) can be written as 
\begin{equation}
\label{Qm Hamiltonian}
H=\frac{1}{2}\left[-\frac{1}{r^2}\frac{\partial}{\partial r}\left(r^2\frac{\partial}{\partial r}\right)+
\frac{1}{r^2}\vJ\!\,^2+\omega^2 r^2\right]\,,
\end{equation} 
where $\vJ$ is just the quantum version of the Poincar\'e integral (\ref{ClassicPoincare}), 
the components of which generate the 
$\mathfrak{su}(2)$ 
symmetry. 
The Dirac quantization condition 
implies that $\nu=eg$ must take a integer or half integer 
value \cite{Mono1,Mono2,Mono2+}.
Using the angular momentum treatment we obtain  
\begin{equation}
\label{so(3) rep1}
{\vJ}\!\,^2\mathcal{Y}_{j}^{j_3}=j(j+1)\mathcal{Y}_{j}^{j_3}\,,\quad
 J_3\mathcal{Y}_{j}^{j_3}=j_3\mathcal{Y}_{j}^{j_3}\,,\quad 
J_\pm\mathcal{Y}_{j}^{j_3}=c_{jj_3}^\pm\mathcal{Y}_{j}^{j_3\pm1}\,, 
\end{equation}
with $J_\pm=J_1\pm i J_2$, 
and 
\begin{equation}
j=|\nu|,|\nu|+1,\ldots\,,\qquad 
 j_3=-j,\ldots,j\,,\qquad
c_{jj_3}^\pm=\sqrt{(j\pm j_3+1)(j\mp j_3)}\,,
\end{equation}
where the indicated values for $j$ correspond to a super-selection rule. 
The case $\nu=0$ corresponds just to the quantum harmonic  isotropic oscillator.
Excluding the zero value for $\nu$, i.e. implying that $\vert \nu\vert$ takes
any nonzero integer or half-integer value,
the first relation in (\ref{so(3) rep1}) automatically 
provides the  necessary inequality  $\vJ^2=j(j+1)>\nu^2$.
The functions $\mathcal{Y}_{j}^{j_3}=\mathcal{Y}_{j}^{j_3}(\theta,\varphi;\nu)$
are the (normalized) monopole harmonics \cite{monoharm,monoharm+,Lochak,Mono1,Mono2,Mono2+},
which are well defined functions if and only if the combination $j\pm\nu$ 
is in $\N_0=\{0, 
1,2,\ldots\}$ 
(see Appendix \ref{armonicosdemonopolo}).

Then, the eigenstates and the spectrum of $H$ are given by 
\begin{align}
\psi_{n,j}^{j_3}(\vr
)&=f_{n,j}(\sqrt{\omega}r)
\mathcal{Y}_j^{j_3}(\theta,\varphi 
)\,,\nonumber
\\
f_{n,j}(x)&=\bigg(\frac{2n!}{\Gamma(n+j+3/2)}\bigg)^{1/2}\omega^{3/4}
\,x^{j}L_{n}^{(j+1/2)}(x^2)\,e^{-x^2/2}\,,\label{wavefunction}\\
 E_{n,j}&=\Big(2n+j+\tfrac{3}{2}\Big)\,\omega\,,\nonumber
\end{align}
where $L_{n}^{(j+1/2)}(y)$ are the generalized Laguerre polynomials. 
The degeneracy of the
energy  level $E_{n,j}$ can be computed by using 
 the property  $E_{n,j}=E_{n+i,j-2i}$
with
 $i\in\{-n,-n+1,\ldots,[(j-\nu)/2]\}$, 
 where $[\,.\,]$ is the integer part,
and 
the fact that  
there are $2(j-2i)+1$ different states with second index  $j-2i$. 
This gives us the sought for degeneracy    
\begin{eqnarray}
\label{ladgen}
\mathfrak{g}(\nu,N)=
\left\{
\begin{array}{ccc}
\tfrac{1}{2}(N+\nu+1)(N-\nu+2)\,, & j-\nu &\text{even 
} \\
\\
\tfrac{1}{2}(N-\nu+1)(N+\nu+2)\,, & j-\nu &\text{odd} 
\end{array}
\right.\,,\qquad N=2n+j\,.
\end{eqnarray}

It is remarkable that
the system possesses  
$2\vert \nu \vert +1$  
degenerate ground states.
The ground states here are not invariant
under the 
action of the
total angular momentum $\vJ$, although the
Hamiltonian  operator commutes with $\vJ$ and hence is
spherically symmetric. Thus we see some analog of
spontaneous breaking of rotational symmetry in
the  
magnetic monopole
background.
This is of course in contrast to the isotropic harmonic oscillator 
in three dimensions which has a unique spherically symmetric
ground state and symmetry algebra $\mathfrak{su}(3)$. 
According to \cite{Vinet} the symmetry algebra for
the system under investigation is
$\mathfrak{su}(2)\oplus\mathfrak{su}(2)$. We do not further
dwell on these interesting aspects of symmetry but 
rather turn to the construction of spectrum generating ladder operators.

Note that the coefficients at radial, $n$,
and angular momentum, $j$, quantum numbers
in the energy eigenvalue  $E_{n,j}=(2n+j+\tfrac{3}{2})\,\omega$ correspond to  
the ratio  $P_a/P_r=l_a/l_r=2$ 
between the classical angular and radial periods 
in the special case $\alpha=\nu^2$ under investigation.
This can be compared with the structure of the
principle quantum number $N=n_r+l+1$
defining the spectrum 
in the quantum model of the hydrogen atom,
where the corresponding classical periods are equal.

\subsection{The algebraic approach}
\label{algebraic approach}
The explicit wave functions in (\ref{wavefunction}) are
specified by the discrete quantum numbers $n$, 
$j$ and $j_3$.
The 
purpose  
of this subsection is to identify the 
 ladder operators for radial, $n$, and 
 angular momentum, $j$, quantum numbers
  (we already have the ladders 
operators for $j_3$),  which are 
based on the conformal  and hidden symmetries of the system. 

In the algebraic approach we do not fix the representation
for the position and momentum operators and thus use
Dirac's ket notation  
for eigenstates.
\vskip0.2cm

\noindent
\emph{Ladder operators for $n$.} Let us first consider 
the quantum version of the $\mathfrak{sl}(2,\R)$ symmetry, 
\begin{equation}
\label{Qsl2r}
[H,\cC]=-2\omega\cC\,,\qquad 
[H,\cC^\dagger]=2\omega\cC^\dagger
\,,\qquad
[\cC,\cC^\dagger]=4\omega H\,,
\end{equation}
where the generators $\cC,\cC^\dagger$ 
are the quantum versions of the integrals (\ref{Cladder}), i.e.
\begin{equation}
\label{ladern}
\cC=
 e^{2i\omega t}\Big(H-\omega^2r^2- \frac{i\omega}{2}(\vr\cdot\vpi+\vpi\cdot\vr)\Big)\,.
\end{equation} 
The time-dependent factors $ e^{2i\omega t}$ 
in $\cC$  and $ e^{-2i\omega t}$ in $\cC^\dagger$
can be omitted without
changing the form of the algebra.
Due to the first two equations in (\ref{Qsl2r}), the scalar nature 
of $\cC$ and $\cC^\dagger$,  and the spectrum 
(\ref{wavefunction}) of the system,
 it is clear that these 
operators change $n$ in $n\pm 1$. 
Then using the relations       
\begin{eqnarray}\label{CCCC}
&\cC\cC^\dagger=H^2+2\omega H-\omega^2\left(\vJ^2-\frac{3}{4}\right)\,,
\qquad
\cC^\dagger\cC=H^2-2\omega H-\omega^2\left(\vJ^2-\frac{3}{4}\right)\,,&
\end{eqnarray} 
one obtains 
\begin{eqnarray}
&
\cC\ket{n,j,j_3}=\omega\, d_{n,j}\ket{n-1,j,j_3}\quad,\quad
\cC^\dagger\ket{n,j,j_3}=\omega\, d_{n+1,j}\ket{n+1,j,j_3}\,,&\\&
d_{n,j}=\sqrt{2n(2n+2j+1)}\,.\label{d-coeff}&
\end{eqnarray}
Rescaling  the $\mathfrak{sl}(2,R)$ generators,  $\mathcal{J}_0=\frac{1}{2\omega} H$, 
$\mathcal{J}_-=\frac{1}{2\omega} \cC$,
$\mathcal{J}_+=\frac{1}{2\omega} \cC^\dagger$,
the conformal algebra takes the form of the $\mathfrak{so}(2,1)$ 
Lorentz algebra 
\be
[\mathcal{J}_0, \mathcal{J}_\pm]=\pm \mathcal{J}_\pm\,,\qquad
[\mathcal{J}_-,\mathcal{J}_+]=
2\mathcal{J}_0
\ee
with Casimir invariant $\mathscr{F}=-\mathcal{J}_0^2+\frac{1}{2}
(\mathcal{J}_+\mathcal{J}_-+\mathcal{J}_-\mathcal{J}_+)$. 
From (\ref{wavefunction}) it follows
that the eigenvalues of $\mathcal{J}_0$ in our case 
are $\mu+n$ with $\mu=\frac{1}{2}(j+\frac{3}{2})$,
and using (\ref{CCCC}), we find that the 
Casimir invariant takes on  eigenstates 
 (\ref{wavefunction}) the value $\mathscr{F}=-\mu(\mu-1)=
 -\frac{1}{4}(j(j+1)-\frac{3}{4})$. 
 If we restore (momentarily)
 Planck's constant and compare with the classical analog (\ref{clasCas}),
 we see that  the last term in parenthesis is equal to
 $-\frac{3}{4}\hbar^2$ and is a quantum correction. 
We conclude that each subspace 
of the Hilbert space of the system characterized  
by the quantum number $j$ carries 
an irreducible unitary infinite-dimensional representation
of the conformal algebra $\mathfrak{sl}(2,\R)$ 
of the discrete type series $D^+_\mu$   
\cite{PlyuSL2R}.  The operators $\cC=2\omega\mathcal{J}_-$ and 
$C^\dagger=2\omega \mathcal{J}_+$
correspond here to the ladder operators 
of  $\mathfrak{sl}(2,\R)$.
\vskip0.1cm

\emph{Ladder operators for $j$.} We introduce the
complex vector operator 
\begin{equation}
\va=\frac{1}{2}(\vb\cross\vJ-\vJ\cross\vb)e^{i\omega t}=(\vb\cross\vJ-i\vb)e^{i\omega t}\,,\qquad
\label{b}
\end{equation}
together with its Hermitian conjugate, where the vector 
$\vb$ has been defined in (\ref{ladders1}).
The vector operator $\va$ is the quantum version of the 
complex classical quantity in (\ref{ladders1}) 
and its components satisfy the relations 
\begin{eqnarray}
&\label{comHaJa}
[H,a_{i}]=-\omega a_i\,,\qquad [J_i,a_{j}]=i\epsilon_{ijk}a_{k}\,,\qquad
[a_{i},a_j]=-i\epsilon_{ijk}\cC J_k\,,
&\\&
\label{comaa}
[a_{i}^\dagger,a_{j}]=-\omega[(2\vJ^2+1-\nu^2)\delta_{ij}-J_{i}J_j)]-iH\epsilon_{ijk}J_k\,,
&
\end{eqnarray} 
with  corresponding relations for the Hermitian conjugate $a_i^\dagger$.
Again, the time-dependent phase $e^{i\omega t}$ can be omitted,
i.e., one can make a change  $\va\rightarrow \va e^{-i\omega t}$ accompanied 
by the analogous omission of the time-dependent phase in the 
generators of the conformal algebra, that does not change the 
form of the commutation
relations (\ref{comHaJa}) and (\ref{comaa}). 

\begin{figure}[H]
\begin{center}
\includegraphics[scale=0.25]{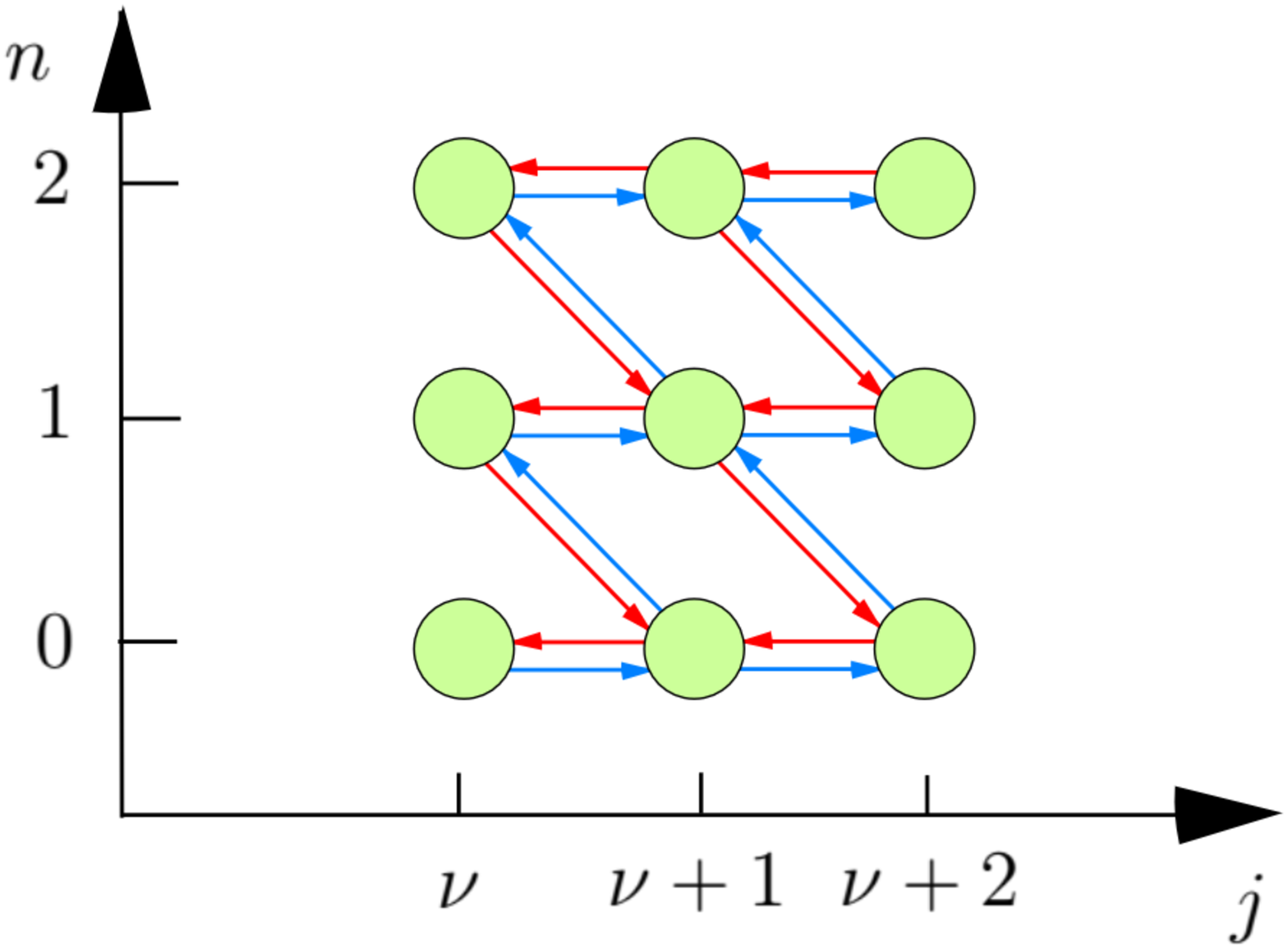}\hskip5mm
\caption{The  
circles 
represent the  first two 
	quantum numbers of the eigenstates
	$\ket{n,j,j_3}$. Red arrows indicate the action of $a_3$ and blue 
	arrows correspond to the action of $a_3^\dagger$. } 
\label{figure2}
\end{center} 
\end{figure}

The action of these operators is computed algebraically in Appendix \ref{derivAB}. 
Here it is sufficient to consider  $a_3$ and $a_3^\dagger$
and their actions on the ket-states
\begin{align}
\label{eta on psi1}
a_3\ket{n,j,j_3}&=A_{n,j,j_3}\ket{n,j-1,j_3}+B_{n,j,j_3}\ket{n-1,j+1,j_3}\,,
\\
a_3^\dagger\ket{n,j,j_3}&=A_{n,j+1,j_3}\ket{n,j+1,j_3}+B_{n+1,j-1,j_3}\ket{n+1,j-1,j_3}\,,
\end{align}
where the squares of the positive coefficients are
\begin{align}
\label{Anlm1}
\big(A_{n,j,j_3}\big)^2&=\omega(2n+2j+1)\,
	\frac{(j^2-j^2_3)(j^2-\nu^2)}{(2j)^2-1}\,,
	\quad
	\big(B_{n,j,j_3}\big)^2&=\frac{2n}{2n+2j+3}\big(A_{n,j+1,j_3}\big)^2\,. 
\end{align}
We see that the operators $a_3$ and $a_3^\dagger$
change the quantum numbers $n$ and $j$, but the result is a 
superposition of the two 
eigenstate 
vectors. Their 
action
is depicted in Fig. \ref{figure2}.
Clearly, it would be preferable to find 
ladder operators that map a given eigenstate
into just one eigenstate
with a  different quantum number $j$ and not a superposition
of eigenstates. To find such operators we introduce 
the non-local operator  
 \begin{eqnarray}
&
\mathscr{J}=\sqrt{\vJ^2+\frac{1}{4}}-\frac{1}{2}\,,\qquad
\mathscr{J}\ket{n,j,j_3}=j\ket{n,j,j_3}\,,
 \end{eqnarray}
and 
construct the operators
\begin{equation}
\mathscr{T}_{\pm}=\omega(\mathscr{J}+\ha)a_3\pm(H-\omega)a_3\mp a_3^\dagger\cC\,
\label{defLadder2}
\end{equation}
together with their Hermitean conjugate. Actually $\mathscr{T}_\pm$ 
and $\mathscr{T}_\pm^\dagger$
are the third components of the vector operators $\mathbfcal{T}_\pm$
and $\mathbfcal{T}_\pm^\dagger$ 
which are given by (\ref{defLadder2}) wherein $a_3$ and $a_3^\dagger$ are replaced by
$\va$ and $\va^\dagger$ on the right hand side.
But in what follows it suffices to consider $\mathscr{T}_\pm$ 
and $\mathscr{T}_\pm^\dagger$
which are ladder operators for the energy,
\be
[H,\mathscr{T}_\pm]=\omega\mathscr{T}_\pm\,,\qquad 
[H,\mathscr{T}_\pm^\dagger]=-\omega\mathscr{T}_\pm^\dagger\,.
\ee
They decrease and increase the angular momentum 
according to
\begin{align}
\label{ActionA-}
\mathscr{T}_+\ket{n,j,j_3}&=\omega(2j+1)A_{n,j,j_3}\ket{n,j-1,j_3}\,,\\
\label{ActionA+}
\mathscr{T}_-\ket{n,j,j_3}&=\omega(2j+3)B_{n,j,j_3}\ket{n-1,j+1,j_3}\,,
\end{align}
and the analogous Hermitian conjugate relations.
These non-local objects were inspired by a similar construction 
presented in 
\cite{QueMosh}
 for the three dimensional isotropic harmonic oscillator.
  
Now one can generate in a simple way all eigenstates of the
commuting observables $H,\vJ^2$ and $J_3$
by acting with the local ladder operators $\cC,\cC^\dagger, J_\pm$ and
with the non-local ladder operators $\mathscr{T}_+,\mathscr{T}_{+}^\dagger$ 
on just  one eigenstate. The same can 
be achieved with local ladder operators when
one uses $a,a^\dagger$ instead of 
$\mathscr{T}_+,\mathscr{T}_+^\dagger$, but then the recursive construction
get more involved, since $a,a^\dagger$ map
into a superposition of eigenstates.

\subsection{The conformal bridge}
\label{Conformal bridge}
Here we show how the generators of the conformal as well as 
the hidden symmetry  of the quantum 
 system (\ref{Qm Hamiltonian}) 
 can be obtained from generators of the corresponding 
 symmetries of the quantum system studied in 
 \cite{N4monopole}. This will be realized  by means 
 of the special non-unitary transformation 
considered recently in \cite{Conformalbridge}  
and identified there as a ``conformal bridge transformation". 
As it will be seen, such a transformation simultaneously 
 generates eigenstates and coherent states of the system   (\ref{Qm Hamiltonian}) 
 from certain states of the quantum 
system considered in 
\cite{N4monopole}.

Similarly to
the classical case, in the limit $\omega\rightarrow 0$ 
the quantum version of the generators (\ref{KyD}) 
has the form
\begin{eqnarray}
&
\label{spin0monopole}
H_0=\frac{1}{2}\left(\vpi^2+\frac{\nu^2}{r^2}\right)=\frac{1}{2}
\left(-\frac{1}{r^2}\frac{\partial}{\partial r}\left(r^2\frac{\partial}{\partial r}\right)+
\frac{1}{r^2}\vJ\!\,^2\right)\,,&
\\&
D_0=\frac{1}{4}(\vr\vpi+\vpi\cdot\vr ) - H_0t\,,\qquad
K_0=\frac{1}{2}
r^2 -Dt-H_0t^2\,.&\label{K0r2}
\end{eqnarray}
They produce the quantum conformal algebra 
\begin{equation}
\label{quantumconformal}
[D_0,H_0]=iH_0\,,\qquad
[D_0,K_0]=-iK_0\,,\qquad
[K_0,H_0]=2iD_0\,. 
\end{equation}
The Hamiltonian $H_0$ is a non-compact generator of the 
conformal algebra $\mathfrak{sl}(2,\R)$ with a continuos spectrum $(0,\infty)$.
In the same limit, the quantum version of the vector integrals $\vI_1$ and 
$\vI_2$ transforms into the vectors 
\begin{equation}
\label{LRL-G}
\vI_1\rightarrow\frac{1}{2}\left(\vpi\cross\vJ-\vJ\cross\vpi\right):=\vV\,,\qquad
\frac{\vI_2}{\omega}\rightarrow\frac{1}{2}\big(\vpi t-\vr)\cross\vJ-\vJ\cross(\vpi t-\vr)\big):=\vG\,,
\end{equation}
which we identify, respectively,  as the  
 Laplace-Runge-Lentz vector and the Galilei boost 
 generator 
for the system $H_0$ \cite{N4monopole} in the Weyl-ordered form.
The commutator relations of the 
vectors $\vV$ and $\vG$ with the generators of the conformal algebra 
are
\begin{eqnarray}
&
[H_0,G_i]=-iV_i\,,\qquad
[K_0,V_i]=iG_i\,,\qquad
[H_0,V_i]=[K_0,G_i]=0&\\&
[D_0,V_i]=\frac{i}{2}V_i\,,\qquad
[D_0,G_i]=-\frac{i}{2}G_i\,.&
\end{eqnarray}
In order to go in the opposite  direction, i.e., to 
recover our system $H$ and its symmetry generators starting from the generators 
(\ref{spin0monopole}), (\ref{K0r2}) and 
 (\ref{LRL-G}), we implement a particular non-unitary transformation. 
The conformal bridge transformation  \cite{Conformalbridge}
corresponds to setting  $t=0$ in the definition 
of the generators above and in the construction of  the operator 
\begin{equation}
\mathfrak{S}=e^{-\omega K_0}e^{\frac{1}{2\omega}H_0}e^{i\ln 2 D_0}\,
\end{equation}
in terms of the conformal symmetry  generators of the system
$H_0$. A similarity transformation generated by  $\mathfrak{S}$ yields 
\begin{eqnarray}
\label{bridge 1}
&
\mathfrak{S}\vJ\mathfrak{S}^{-1}=\vJ\,,\qquad
\mathfrak{S}\vV\mathfrak{S}^{-1}=\va\,,\qquad
\mathfrak{S}\omega\vG\mathfrak{S}^{-1}=-i\va^\dagger\,,&\\
&
\label{bridge 2}
\mathfrak{S}(H_0)\mathfrak{S}^{-1}=\frac{1}{2}\cC\,\qquad
\mathfrak{S}(2i\omega D_0)\mathfrak{S}^{-1}=H
\,,\qquad
\mathfrak{S}(\omega^2 K_0)\mathfrak{S}^{-1}=-\frac{1}{2}\cC^\dagger\,,
&
\end{eqnarray}
where $H=H_0+\omega^2K_0$ is the quantum Hamiltonian (\ref{Qm Hamiltonian}). 
The explicit time dependence can  be recovered then  by applying a
unitary transformation with the evolution operator 
$U=e^{-iHt}$.
 
In correspondence with the second relation 
in (\ref{bridge 2}),  the 
wave functions (\ref{wavefunction}) can be generated, up to a normalization, by applying
the non-unitary operator  $\mathfrak{S}$
to  the  formal  eigenstates of the first order differential operator $2i\omega D_0$ defined 
by the  equation 
$2i\omega D_0\chi_{n,j}^{j_3}=\omega(2n+j+3/2)\chi_{n,j}^{j_3}$. The latter are given by
\begin{equation}
\chi_{n,j}^{j_3}(r,\theta,\phi)=r^{j+2n}\mathcal{Y}_{j}^{j_3}(\theta,\phi)\,.
\end{equation} 
To see that $\mathfrak{S}$ has these properties, we 
consider the following relation which can  be proved by induction, 
\begin{equation}
\label{JOrdan}
(H_0)^{\ell}\chi_{n,j}^{j_3}(\vr)=
\frac{(-2)^\ell\Gamma(n+1)\Gamma(n+j+3/2)}{\Gamma(n+1-\ell)\Gamma(n+j+3/2-\ell)}\chi_{n-\ell,j}^{j_3}\,,\qquad
\ell=1,2,\ldots\,.
\end{equation}
It vanishes for $\ell>n$
 due to the pole of the Gamma function in the denominator. 
Therefore,  
$\chi_{0,j}^{j_3}$ can be interpreted 
as the zero-energy eigenstate of $H_0$, and $\chi_{n,j}^{j_3}$, $n=1,2,\ldots$,  are its 
Jordan states corresponding to the same zero energy \cite{InzPly4}. 
Decomposing the operator $\exp(\frac{1}{2\omega}H_0)$ entering
the definition of $\mathfrak{S}$
and using equation (\ref{JOrdan}) one obtains 
\begin{eqnarray}
&\mathfrak{S}\chi_{n,j}^{j_3}=\frac{(-1)^n}{\sqrt{2}}\left(\frac{2}{\omega}\right)^{n+\frac{j}{2}+\frac{3}{4}}
\left[n! \Gamma(n+j+3/2)\right]^{\frac{1}{2}}\psi_{n,j}^{j_3}\,. &
\end{eqnarray}
On the other hand, one can show that  solutions 
of the equation $H_0\phi^{j_3}_{j}=\frac{1}{2}\kappa^2\phi_j^{j_3}$
are mapped into coherent states of $H$. The explicit form of these 
solutions is given by
\begin{eqnarray}
\label{Jstates}
&\phi_j^{j3}(\vr;\kappa)=\frac{1}{\sqrt{r}}J_{j+\frac{1}{2}}(\kappa r)\mathcal{Y}_{j}^{j_3}
=\sum_{n=0}^{\infty}\frac{(-1)^{n}(\kappa/2)^{2n+j+1/2}}{n!\Gamma(n+j+3/2)}\chi_{n,j}^{j_3}(\vr)\,,&
\end{eqnarray}
where in the first equality 
$J_{j+\frac{1}{2}}(y)$ is the Bessel function of the first kind, 
and in the second one we show that the solution can be expressed as 
a series expansion in terms of 
the Jordan states (\ref{JOrdan}). The explicit action of 
$\mathfrak{S}$ on (\ref{Jstates}) gives us 
\begin{eqnarray}
\label{coherent}
&\zeta^{j_3}_{j}(\vr;\kappa)=
N\mathfrak{S}\phi_j^{\,j3}(\vr;\frac{\kappa}{\sqrt{2}})=\sqrt{2}Ne^{-\frac{\omega 
x^2}{2}+\frac{\kappa^2}{4\omega}}\phi_j^{\,j3}(\vr;\kappa)\\&=
\frac{N}{\omega^{1/2}}\sum_{n=0}^{\infty} \frac{1}
{\sqrt{n!\Gamma(n+j+\frac{3}{2})}}\left(\frac{\kappa}{2\sqrt{\omega}}\right)^{2n+j+
1/2}\psi_{n,j}^{j_3}(\vr)\,,
&
\end{eqnarray}
where $N$ stand for a normalization constant. Using the 
series expansion in terms of $\psi_{n,j}^{j_3}$,  one finds 
$N^2=\sqrt{\omega}/(I_{j+\frac{1}{2}}(\frac{|\kappa|^2}{2\omega}))$, 
where  $I_{j+\frac{1}{2}}(z)$ is the modified Bessel function of the first kind,
and we have put  the modulus in its argument 
because $\kappa$ admits an analytic extension for complex values, as is usual 
for coherent states. 
Also, this expansion helps us to find the time evolution of these functions 
generated by  $H$ as follows,
\begin{equation}
\zeta^{j_3}_{j}(\vr,t;\kappa)=
e^{-iHt}\zeta^{j_3}_{j}(\vr;\kappa)=
e^{-i\frac{\omega}{2} t}\zeta^{j_3}_{j}
(\vr;\kappa e^{-\frac{i\omega t}{2}})\,.
\end{equation} 
At the same time, the first equation in (\ref{bridge 2})
 shows us that $\zeta^{j_3}_{j}(\vr,t;\kappa)$  are eigenstates 
of $\mathcal{C}$ with eigenvalue $-\frac{1}{2}\kappa^2 e^{-i\omega t}$, i.e. 
they are indeed the coherent states 
of the system 
corresponding to the conformal algebra
(\ref{Qsl2r}) \cite{Perelomov}.

\section{A charge-monopole superconformal model}
\label{SUSYsection}

In this section we admit an additional contribution to
the
Hamiltonian (\ref{Qm Hamiltonian})
due to  the spin degrees of freedom of the  particle. 
The
additional term describes
a strong long-range spin-orbit coupling and gives rise 
to an exactly solvable supersymmetric extension
of the three-dimensional system studied in previous sections.
After introducing the system in the first subsection we
analyze its  peculiar properties. 
In particular, we will 
find that
some energy values  are infinitely degenerate and 
others are not.  
In the second subsection, we show 
how one can extend the model to a supersymmetric system 
by means of the factorization method, and  by using this we construct
an explicit realization of the $\mathfrak{osp}(2\vert 2)$ superconformal symmetry. 
In the third subsection we show that two different superconformal 
extensions of the one-dimensional AFF model with unbroken and spontaneously broken 
phases of $\mathcal{N}=2$ Poincar\'e supersymmetry
can be obtained by  reduction from 
 our three-dimensional $\mathfrak{osp}(2\vert 2)$ superconformal system.

\subsection{Spin-orbit coupling model}
Let us consider the following two Hamiltonians with strong spin-orbit coupling  
\begin{equation}
\label{spinorbitH}
H_{\pm\omega}= \frac{1}{2}\left(\vpi^2 +\omega^2 r^2+\frac{\nu^2}{r^2}\right)\pm\omega\vsigma\cdot\vJ
=H\pm\omega\vsigma\cdot\vJ\,.
\end{equation}
The Hamiltonians $H_{\pm\omega}$ are similar to those which appear
as subsystems of the non-relativistic limit of the supersymmetric 
Dirac oscillator discussed in \cite{DiracOs,RevLett}. Thus
the eigenvalue problems can be solved similarly as in those 
references, but the usual spherical harmonics
are replaced by the monopole harmonics. Actually,
if we would choose a spin-orbit coupling $\omega'\vsigma\cdot\vJ$
with $0\leq \omega<\omega'$, then the spectra of both
Hamiltonians would be unbounded 
from
below. 
On the other hand, for
$0\leq \omega'<\omega$ all energies will have finite
degeneracy. Only in the very particular case $\omega'=\omega$, which we 
consider here, the spectra are bounded from below and
half of the energies have a finite degeneracy whereas
the other half have infinite degeneracy. This 
reminds us  the BPS-limits in 
field theory, where different interactions balance and
supersymmetry is observed.

The operators $H$ and $\vsigma\cdot\vJ$ commute and as a consequence
$H_{\pm \omega}$ commute with the ``total angular momentum"
\begin{equation}
\vK=\vJ+\vs=\vJ+\tfrac{1}{2}\vsigma\,,\qquad
[K_i,K_j]=i\epsilon_{ijk}K_k\,.
\end{equation}
The possible eigenvalues of $\vK^2$ are $k(k+1)$.
It is well-known how to construct the simultaneous eigenstates
of $\vK^2$ and $K_3$:
\begin{equation}
\ket{n,k,k_3,\pm}
=\sum_{m_s}C^{kk_3}_{jj_3\has m_s} \ket{n,j,j_3}
\otimes\ket{\ha,m_s}_{k=j\pm\has}\,,
\label{eigenHpm}
\end{equation}
where the Clebsch-Gordan
 coefficients 
\begin{equation}
C^{kk_3}_{jj_3\has m_s}=\bra{j,j_3,\tfrac{1}{2},m_s}k,k_3\rangle\,
\end{equation}
on the right hand
side are nonzero only if $j_3+m_s=k_3$ and 
if the triangle-rule holds, which means that
the total angular momentum $k$ is either $j+\frac{1}{2}$ 
or $j-\frac{1}{2}$. In the first case the eigenstates
of the total angular momentum are
denoted by $\ket{\dots,k,k_3,+}$ and in the second case
by $\ket{\dots,k,k_3,-}$.
The sums (\ref{eigenHpm}) contain just two terms, since the eigenvalue $m_s$ 
of the third spin-component 
$s_3=\frac{1}{2}\sigma_3$ 
is either $\frac{1}{2}$ 
or $-\frac{1}{2}$.
Note that in the coordinate representation  
the wavefunctions corresponding
to these kets are given
in (\ref{wavefunction}), i.e.
\begin{eqnarray}
&
\label{Wspin+-}
\bra{\vr}\ket{n,k,k_3,\pm}=f_{n,j}(\sqrt{\omega}r)\bra{\vn}\ket{k,k_3,\pm}\,,&\\&
\label{Omega}
\bra{\vn}\ket{k,k_3,\pm}=
\frac{1}{\sqrt{2k+1\mp 1}}
\left(\begin{array}{cc}
\pm \sqrt{k\pm k_3+(1\mp 1)/2}\,\mathcal{Y}_{k\mp 1/2}^{k_3-1/2}(\theta,\varphi;\nu)\\
\sqrt{k\mp k_3 +(1\mp 1)/2}\,\mathcal{Y}_{k\mp 1/2}^{k_3+1/2}(\theta,\varphi;\nu)
\end{array}\right):=\Omega_{k}^{k_3\,\pm}\,.
&
\end{eqnarray}
The wavefunctions 
(\ref{Wspin+-})
contain the monopole harmonics
and generalized Laguerre polynomials.
If $\nu=eg$ is integer-valued then $j$ is a non-negative
integer and $k$ a positive half-integer. If $eg$ is half-integer, then
$j$ is a positive half-integer and $k$ is in $\N_0$.

The vector 
in (\ref{eigenHpm}) is a simultaneous eigenstate of $\vJ^2$ 
with eigenvalue $j(j+1)$, of $\vK^2$ with eigenvalue $k(k+1)$, of $H$ with eigenvalue
$(2n+j+\frac{3}{2})\omega$,  where $j=k\mp 1/2$, 
and finally of the operator $\vsigma\cdot\vJ$:
\begin{equation}
\vsigma\cdot\vJ\ket{n,k,k_3,\pm}=\big(\pm(k+\tfrac{1}{2})-1\big)
\ket{n,k,k_3,\pm}\,.
\end{equation}
As a consequence the action of the Hamiltonians in (\ref{spinorbitH}) 
on these states is 
\begin{align}
H_{+\omega}\ket{n,k,k_3,\pm}&=\omega\left(2n+k+\tfrac{1}{2} \pm k \right)
\ket{n,k,k_3,\pm}\,,\label{Hket+}\\
H_{-\omega}\ket{n,k,k_3,\pm}&=\omega\left(2n+k+\tfrac{5}{2} \mp(k+1)\right)
\ket{n,k,k_3,\pm}\,.
\label{Hket-}
\end{align} 
We see that the discrete eigenvalues of both
Hamiltonians $H_{\pm\omega}$ fall into two families:
in one
family all energies are  infinitely degenerate
and in the other family they all have finite degeneracy
(due to their dependence on the quantum number $k$). 
More explicitly, for $k=j\mp\frac{1}{2}$ the eigenvalues 
of $H_{\mp\omega}$ have infinite degeneracy
and for $k=j\pm\frac{1}{2}$ they  have finite 
degeneracy $\mathfrak{g}(N,\nu)=N^2-\nu^2$,  where $N=n+j+1$.
A similar peculiar behavior is observed in the Dirac 
oscillator spectrum \cite{DiracOs}.

Operators $K_\pm=K_1\pm iK_2$ are the ladder operators
for the magnetic quantum number $k_3$. 
The ladder operators for the radial quantum number 
are 
given in (\ref{ladern}), 
and their 
action
on the simultaneous eigenstates 
reads
\begin{align}
\cC\ket{n,k,k_3,\pm}&=\omega d_{n,j}\ket{n-1,k,k_3,\pm}\,,\\
\cC^\dagger\ket{n,k,k_3,\pm}&=\omega d_{n+1,j}\ket{n+1,k,k_3,\pm}\,,
\end{align}
with coefficients defined in (\ref{d-coeff}).
Thus, as for the spin-zero particle system in monopole background, 
we can
easily construct local ladder operators for $n$ and $k_3$.
But again, finding ladder operators for $k$ is
more difficult.
One way to proceed is to follow the ideas employed
for the Dirac oscillator in \cite{QueMosh,Quesne}.
First we decompose the total Hilbert space in two subspaces,
$\mathscr{H}=\mathscr{H}^{(+)}\oplus \mathscr{H}^{(-)}$, where 
each
$\mathscr{H}^{(\pm)}$ is spanned by the states 
$\ket{n,k,k_3,\pm}$.
Actually we can construct non-local operators 
which project orthonormally onto these subspaces,
\begin{align}
\mathscr{P}_{+}&=\ha+\sqrt{\vK^2+\tfrac{1}{4}}-
\sqrt{\vJ^2+\tfrac{1}{4}}\,,\\
\mathscr{P}_{-}&=\ha-\sqrt{\vK^2+\tfrac{1}{4}}+
\sqrt{\vJ^2+\tfrac{1}{4}}\,,
\end{align}
and 
reproduce or annihilate the eigenstates,
\begin{equation}
\mathscr{P}_{(\pm)}\big\vert_{\mathscr{H}^{(\pm)}}
=\id\big\vert_{\mathscr{H}^{(\pm)}}\,,\qquad
\mathscr{P}_{(\pm)}\big\vert_{\mathscr{H}^{(\mp)}}=0\,.
\end{equation} 
In next step we introduce the operators  
\begin{equation}
\label{PAP+}
\mathcal{A}_{\pm}=\mathscr{P}_\pm \mathscr{T}_\pm\mathscr{P}_\pm\,,
\end{equation}
where the non-local $\mathscr{T}_\pm$ have been
defined in (\ref{defLadder2}). 
The presence of the projectors will ensure that 
$\mathcal{A}_\pm$ only
acts 
on eigenstates in $\mathscr{H}^{(\pm)}$, and 
its action on these eigenstates  
can be computed straightforwardly using the relations (\ref{ActionA-}) 
and (\ref{ActionA+}):
\begin{align}
\label{nonlocalAaction}
&\mathcal{A}_+ \ket{n,k,k_3,+}=
(k-1)\sqrt{n+k}\,\Lambda_{k,k_3,j}\, 
\ket{n,k-1,k_3,+}\,, 
\\
&\mathcal{A}_{-} \ket{n,k-1,k_3,-}=
(k+1)\sqrt{n}\,\Lambda_{k,k_3,j}
\, 
\ket{n-1,k,k_3,-}\,,
\end{align}
with 
\[
\Lambda_{k,k_3,j}=\frac{\omega^{3/2}}{k}\sqrt{2(k^2-k_3^2)(j^2-\nu^2)}\,.
\]
These relations mean that the operators $\mathcal{A}_\pm$ 
and their adjoint act as ladder operators for the quantum 
number $k$. Together with operators $K_\pm,\cC,\cC^\dagger$
they generate all eigenstates in the full Hilbert space 
from just two eigenstates, one from each subspace 
$\mathscr{H}^{(\pm)}$.

\subsection{The $\mathfrak{osp}(2\vert 2)$ superconformal extension}
\label{osp22 extension}
In this subsection we construct and analyze  supersymmetric partners
of the Hamiltonians $H_{\pm \omega}$ by introducing 
 factorizing operators. 
From these we obtain two $\mathcal{N}=2$ super-Poincar\'e quantum systems
which are related to each other by a common integral of motion
which generates an $R$-symmetry.
Supplementing the supercharges of one of these systems  by supercharges of another, 
we extend the $\mathcal{N}=2$ super-Poincar\'e  symmetry up to 
the $\mathfrak{osp}(2\vert 2)$ superconformal symmetry 
realized by a three-dimensional system of 
spin-1/2 particle in a monopole background.

Consider 
the  first-order scalar operators 
\begin{equation}
\label{intertwiningQ}
\Theta=i\vsigma\cdot\vb-\frac{1}{\sqrt{2}}\frac{\nu}{r}\,,\qquad 
\Xi=i\vsigma\cdot\vb^\dagger-\frac{1}{\sqrt{2}}\frac{\nu}{r}\,,
\end{equation}
and their adjoint $\Theta^\dagger$ and $\Xi^\dagger$.
The products of these operators with their adjoint are
\begin{equation}
H_{[1]}:=
\Theta\Theta^\dagger=H_{+\omega}+\tfrac{3}{2}\omega\,,\qquad
\breve{H}_{[1]}:=
\Xi\Xi^\dagger=H_{-\omega}-\tfrac{3}{2}\omega\,,
\label{H0}
\end{equation}
where $H_{\pm\omega}$ are given in (\ref{spinorbitH}).
The associated superpartners take the form
\begin{align}
\label{H0a}
H_{[0]}&:=
\Theta^\dagger \Theta=\breve{H}_{[1]}-\nu\left(\tfrac{1}{r^2}+2\omega \right)\sigma_r
\,,
\\
\label{H0b}
\breve{H}_{[0]}&:=\Xi^\dagger \Xi
=H_{[1]}-\nu\left(\tfrac{1}{r^2}-2\omega \right)\sigma_r \,,
\end{align} 
wherein the projection of $\vsigma$ to the normal
unit vector appears,
\begin{equation}
\label{sigman}
\sigma_r=\vn\cdot\vsigma=\left(\begin{array}{cc}
\cos\theta & e^{-i\varphi}\sin\theta\\
e^{i\varphi}\sin\theta & -\cos\theta
\end{array}\right)\,.
\end{equation}
The first order operators satisfy the intertwining relations 
\begin{eqnarray}
&
\Theta H_{[0]}=H_{[1]}\Theta\,,\qquad 
\Theta^\dagger H_{[1]}=H_{[0]}\Theta^\dagger\,,  
&\\&
\Xi\breve{H} _{[0]}=\breve{H}_{[0]}\Xi\,,\qquad
\Xi^\dagger \breve{H}_{[1]}=\breve{H}_{[0]}\Xi^\dagger\,. 
&
\end{eqnarray} 
The eigenstates of $H_{[0]}$ can be obtained by 
acting with $\Theta^\dagger$ on the eigenstates of $H_{[1]}$ or equivalently of $H_{+\omega}$.
They are given in (\ref{eigenHpm}).
For computing the action of the intertwining operators
we use the following representation
\begin{align}
\label{Qnu}
\Theta^\dagger
=\frac{\sigma_r}{\sqrt{2}}\left(-
\frac{1}{r}\frac{\partial}{\partial r} r+\omega r+
\frac{1+\vsigma\cdot\vJ}{r}\right)\,.
\end{align}
As a result we obtain the relations 
\begin{align}
\Theta^\dagger\ket{n,k,k_3,\pm}&=\pm\sqrt{2\omega(n+1+\beta_{\pm}k)}
\,\Vert n+\beta_{\mp},k,k_3,\pm\rangle\,,\quad \beta_\pm=\tfrac{1}{2}(1\pm 1)\,,\label{qdonpsi}
\end{align}
where in coordinate representation the normalized spinors $\Vert n,k,k_3,\pm\rangle$ 
on the right hand 
side
have the explicit form
\begin{align}
\langle\vr\Vert n,k,k_3,\pm\rangle&=f_{n,j\pm1}\sigma_r\Omega_{k}^{k_3\,\pm}\,,
\label{spinors2a}\end{align}
where $\Omega_{k}^{k_3\,\pm}$ are given in (\ref{Omega}). 
With the help of functional relations between
different generalized Laguerre polynomials (see Appendix \ref{Laguerrerelations}) one proves that
\begin{align}
\Theta\Vert n,k,k_3,\pm\rangle&=\pm
\sqrt{2\omega(n+\beta_{\pm}(k+1))}\,\ket{n-\beta_{\mp},k,k_3,\pm}\,,
\label{Qonphi+}
\end{align} 
and by acting with the operator $\Theta^\dagger$ 
on these relations we get the eigenvalue equations
\begin{align}
H_{[0]}\Vert n,k,k_3,\pm\rangle&=2\omega(n+\beta_{\pm}(k+1))\Vert n,k,k_3,\pm\rangle\,.
\end{align} 
Note that the states $\Vert n,k,k_3,-\rangle$ are zero-modes of
$H_{[0]}$ since they are annihilated by $\Theta$. 

The eigenstates of $\breve{H}_{[0]}$ can be
determined in an analogous way by
acting with the operator
\begin{align}
\label{Wsigmaform}
\Xi^\dagger
=\frac{\sigma_r}{\sqrt{2}}\left(-
\frac{1}{r}\frac{\partial}{\partial r} r-\omega r+
\frac{1}{r}(1+\vsigma\cdot\vJ)
\right)\,
\end{align}
on the spinors (\ref{Wspin+-}) and with its adjoint
$\Xi$ on the spinors  (\ref{spinors2a}),  
that
results in the mappings  
\begin{align}
\Xi^\dagger\ket{n,k,k_3,\pm}&=\pm\sqrt{2\omega (n+\beta_{\mp}(k+1)}\,
\Vert n-\beta_\pm,k,k_3,\pm\rangle\,,
\label{W+onpsi1} \\
\Xi\,\Vert n,k,k_3,\pm\rangle
&=\pm\sqrt{2\omega (n+1+\beta_\mp k)}\,\,\ket{n+\beta_{\pm},k,k_3,\pm}\,.
\label{W+onphi2}
\end{align}
Note that $\Xi^\dagger$ as well as $\breve{H}_{[1]}$ annihilate the set  
of states $\Vert 0,k,k_3,+\rangle$.
 
Finally, acting with $\Xi^\dagger$ on the states in
(\ref{W+onphi2}), we solve the eigenvalues problem
for $\breve{H}_{[0]}$:
\begin{align}
\breve{H}_{[0]}
\,\Vert n,k,k_3,\pm\rangle&=2\omega(n+1+k\beta_{\mp})\,\Vert n,k,k_3,\pm\rangle\,.
\label{eigenbreve}
\end{align} 
Having at hand the eigenstates $\Vert n,k,k_3,\pm\rangle$,
one may find spectrum generating ladder operators. In 
this context  equations (\ref{qdonpsi}), (\ref{Qonphi+}), (\ref{W+onpsi1}) and (\ref{W+onphi2}) 
can be used to construct such operators 
for the quantum number $n$. They read
\begin{align}
\tilde{\mathcal{C}}=\Xi^\dagger \Theta\,,\qquad
\tilde{\mathcal{C}}^\dagger=\Theta^\dagger \Xi\,,
\end{align} 
and act on the eigenvectors  $\Vert \dots\rangle$
as 
follows:
\begin{align}
\tilde{\mathcal{C}}^\dagger\,\Vert n,k,k_3,\pm\rangle&=2\omega d_{n+1,j\pm 1}
\,
\Vert n+1,k,k_3,\pm\rangle\,,\nonumber\\
\tilde{\mathcal{C}}\,\Vert n,k,k_3,\pm\rangle&=2\omega d_{n,j\pm 1}
\,
\Vert n-1,k,k_3,\pm\rangle\,.
\label{W+Q}
\end{align} 
Actually, the first order operators $\Theta$ and $\Xi^\dagger$ 
factorize the earlier considered second order 
ladder operator (\ref{ladern}) according to $\cC=\Theta\Xi^\dagger$.

Having constructed lowering and 
raising operators for $n$, 
we are still missing ladder operators for $k$ and $k_3$. For the 
latter we may of course use $K_\pm$, since
$\Theta$, $\Xi$ and their adjoint 
are scalar operators with respect to $\vK$. But once more,
for the angular momentum quantum number $k$ we can introduce
non-local 
``dressed" operators
%\begin{align}
\begin{eqnarray}
\label{nonlocaldressed}
&\tilde{\mathcal{A}}_-=\Theta\sqrt{\frac{1}{H_{[1]}}}\mathcal{A}_-\sqrt{\frac{1}{H_{[1]}}}\Theta^\dagger\,,\qquad
\tilde{\mathcal{A}}_+=\Xi\sqrt{\frac{1}{\breve{H}_{[1]}}}\mathcal{A}_+\sqrt{\frac{1}{\breve{H}_{[1]}}}\Xi^\dagger\,,&\qquad
\end{eqnarray}
%\end{align}
and their  
adjoint
operators, where $\mathcal{A}_\pm$ 
have been given in (\ref{PAP+}). The operators $\tilde{\mathcal{A}}_\pm$
are the analogs to $\mathcal{A}_\pm$ for the vectors $\Vert n,k,k_3,\pm\rangle$, 
as we see in equations 
\begin{align}
\label{nonlocalAaction2}
&\tilde{\mathcal{A}}_+ \Vert n,k,k_3,+\rangle=
(k-1)\sqrt{n+k}\,\Lambda_{k,k_3,j}\, \Vert n,k-1,k_3,+\rangle \,, 
\\
&\tilde{\mathcal{A}}_{-}\Vert n,k-1,k_3,-\rangle=
(k+1)\sqrt{n}\,\Lambda_{k,k_3,j}
\, \Vert n-1,k,k_3,-\rangle\,.
\end{align}

In a final step we combine the four $2\times 2$ matrix
Hamiltonians introduced above into two 
$4\times 4$ matrix super-Hamiltonians  
 as follows:
\begin{eqnarray}
\label{superH}
\mathcal{H}=\left(\begin{array}{cc}
H_{[1]} & 0\\
0 & H_{[0]}
\end{array}\right)\,,\qquad
\breve{\mathcal{H}}=\left(\begin{array}{cc}
\breve{H}_{[1]} & 0\\
0 & \breve{H}_{[0]}
\end{array}\right)\,.
\end{eqnarray}
In the limit $\nu\rightarrow 0$ they turn into
different versions  of the Dirac oscillator in the non-relativistic limit, see \cite{DiracOs}.
Both operators commute with 
the $\Z_2$-grading operator $\Gamma=\sigma_3\otimes \mathbb{I}_{2\times  2}$, 
$[\Gamma,\mathcal{H}]=[\Gamma,\breve{\mathcal{H}}]=0$,
and 
their difference is the (bosonic) integral of motion
\begin{equation}
\mathcal{R}=\frac{1}{2\omega}(\mathcal{H}-\breve{\mathcal{H}})=
(\vJ\cdot \vsigma +\tfrac{3}{2})\Gamma-2\nu\sigma_r\Pi_-= 
\left(\begin{array}{cc}
\vsigma\cdot\vJ+\frac{3}{2} & 0\\
0&-(\vsigma\cdot\vJ+2\nu\sigma_r+\frac{3}{2}) \\
\end{array}\right),
\label{rwisym}
\end{equation} 
where  $\Pi_-$ is a projector,  
\be\label{Pi+-}
\Pi_\pm=\tfrac{1}{2}(1\pm\Gamma)\,.
\ee
In the fermionic sectors of the systems $\mathcal{H}$ and $\breve{\mathcal{H}}$
we have the nilpotent operators 
\begin{eqnarray}
\label{QyW}
&
\cQ=\left(\begin{array}{cc}
0 & \Theta\\
0 & 0
\end{array}\right)\,,\qquad
\cW^\dagger=\left(\begin{array}{cc}
0 & \Xi \\
0 & 0
\end{array}\right)\,,&
\end{eqnarray} 
$\{\Gamma,\mathcal{Q}\}=\{\Gamma,\mathcal{W}\}
=0$, and their adjoint operators. 
Each of these generate
an $\mathcal{N}=2$ Poincar\'e  superalgebra  
\begin{eqnarray}
\label{N2Poincaresuper}
&
[\mathcal{H},\mathcal{Q}]= 
[\mathcal{H},\mathcal{Q}^\dagger]=
\{\mathcal{Q},\mathcal{Q}\}=
\{\mathcal{Q}^\dagger,\mathcal{Q}^\dagger\}
=0\,,\qquad
\{\mathcal{Q},\mathcal{Q}^\dagger\}=\mathcal{H}\,,&\\&
[\breve{\mathcal{H}},\mathcal{W}]=
[\breve{\mathcal{H}},\mathcal{W}^\dagger] 
=\{\mathcal{W},\mathcal{W}\}=
\{\mathcal{W}^\dagger,\mathcal{W}^\dagger\}=0\,,\qquad
\{\mathcal{W},\mathcal{W}^\dagger\}=\breve{\mathcal{H}}\,.
&
\end{eqnarray}
The even integral $\mathcal{R}$ in (\ref{rwisym})
generates an $R$-symmetry
for both systems, and satisfies the relations 
(for details see 
 Appendix \ref{Apcommutator}),
\begin{equation}
[\Gamma,\mathcal{R}]=0\,,\qquad
[\mathcal{R},\mathcal{Q}]=\mathcal{Q}\,,\qquad
[\mathcal{R},\mathcal{W}]=-\mathcal{W}\,,\qquad \text{\text{h.c.}}\,,
\end{equation} 
where \text{h.c.} corresponds to Hermitian conjugate relations.
Having in mind 
that $\mathcal{H}$ and $\breve{\mathcal{H}}$
can be diagonalized 
simultaneously, from now on 
we treat  $\mathcal{H}$ as the Hamiltonian of the super-extended system
and $\breve{\mathcal{H}}=\mathcal{H}-2\omega\mathcal{R}$ as its integral.
Then,  
by anti-commuting $\cQ$ and $\cW$ we obtain the bosonic generator 
\begin{equation}
\mathcal{G}=\{\mathcal{W},\mathcal{Q}\}=\left(\begin{array}{cc}
\mathcal{C} & 0\\
0 & \tilde{\mathcal{C}}
\end{array}\right)\,, \qquad [\Gamma,\mathcal{G}]=0\,,\qquad \text{\text{h.c.}}\,,
\end{equation}
composed from 
 the ladder operators of  
 sub-systems $H_{[1]}$ and $H_{[0]}$ 
 of our system $\mathcal{H}$. 

Taking together, these  scalar generators with respect to  
\begin{equation}
 \mathcal{K}_i=
 \left(\begin{array}{cc}
K_i & 0\\
0 & K_i
\end{array}\right)\,, \qquad i=1,2,3\,,
\end{equation}
obey the superalgebraic  relations 
\begin{eqnarray}
\label{osp(2,2),1}
&
[\mathcal{H},\mathcal{G}]=-2\omega \mathcal{G}\,,\qquad
[ \mathcal{G},\mathcal{G}^\dagger]=4\omega(\mathcal{H}-\omega\mathcal{R})\,,&\\&
[\mathcal{H},\mathcal{W}]=-2\omega\mathcal{W}\,,\qquad
[\mathcal{R},\mathcal{W}]=-\mathcal{W}\,,\qquad
[\mathcal{R},\mathcal{Q}]=\mathcal{Q}\,,\qquad
&\\&
[\mathcal{G},\mathcal{Q}^\dagger]=-2\omega\mathcal{W}\,,\qquad
[\mathcal{G},\mathcal{W}^\dagger]= 2\omega \mathcal{Q}\,,
&\\&
\{\mathcal{Q},\mathcal{Q}^\dagger\}=\mathcal{H}\,,\qquad
\{\mathcal{W},\mathcal{W}^\dagger\}=(\mathcal{H}-2\omega\mathcal{R})\,,
&\\&
\label{osp(2,2),7}
\{\mathcal{Q},\mathcal{W}\}=\mathcal{G}\,,
&
\end{eqnarray}
supplemented by the adjoint  relations. 
The (anti)-commutators not displayed here do vanish.
This superalgebra is identified as 
the  $\mathfrak{osp}(2\vert  2)$ superconformal symmetry
which appears in systems like one-dimensional  harmonic 
super-oscillator or the  superconformal mechanics model with a confining term 
\cite{InzPly,CIP,LM2,InzPly4}. Therefore our construction maybe
considered as generalization of the three-dimensional versions
of these systems  in the  monopole background.

The relations  (\ref{osp(2,2),1})-(\ref{osp(2,2),7})
are invariant under the automorphism $\mathcal{H}\rightarrow\breve{\mathcal{H}}$,
$\mathcal{R}\rightarrow\mathcal{R}$, $\mathcal{Q}\leftrightarrow \mathcal{W}$, $\mathcal{G}\rightarrow\mathcal{G}$ and
h.c.,
which amount to using $\breve{\mathcal{H}}$ instead of
$\mathcal{H}$, as super-Hamiltonian. 
The common eigenstates of $\mathcal{H}$, $\mathcal{R}$, $\Gamma$, $\mathcal{K}_3$ and 
$\mbfgr{\mathcal{K}}^2$ are given by      
\begin{eqnarray}
\label{supervectors}
\ket{n,k,k_3,\pm,1}=\left(\begin{array}{cc}
\ket{n,k,k_3,\pm}\\
0
\end{array}\right)
\,,\qquad
\ket{n,k,k_3,\pm,-1}=\left(\begin{array}{cc}
0\\
\Vert n,k,k_3,\pm\rangle
\end{array}\right)\,,
\end{eqnarray}
which satisfy the eigenvalue equations 
\begin{align}
\label{spectalEq}
\mathcal{H}\ket{n,k,k_3,\pm,\gamma}&=2\omega\big(n+\tfrac{1}{2}(1+\gamma)+\beta_\pm(k+
\tfrac{1}{2}(1-\gamma))\big)\ket{n,k,k_3,\pm, \gamma 
}\,, \\
\Gamma \ket{n,k,k_3,\pm,\gamma}&=
\gamma \ket{n,k,k_3,\pm,\gamma} \,,\qquad \gamma=\pm 1
\,, \\
\label{REq}
\mathcal{R} \ket{n,k,k_3,\pm,\gamma}&= 
[\pm(k+\tfrac{1}{2})+\tfrac{\gamma}{2}] \ket{n,k,k_3,\pm,\gamma}\,,
\\
\label{Konsusy}
\mbfgr{\mathcal{K}}^2\ket{n,k,k_3,\pm,\gamma}&= 
k(k+1) \ket{n,k,k_3,\pm,\gamma}\,,\\
\mathcal{K}_3\ket{n,k,k_3,\pm,\gamma}&=k_3\ket{n,k,k_3,\pm,\gamma}\,.
\end{align}  
The operators $\mathcal{Q}$ and $\mathcal{Q}^\dagger$ ($\mathcal{W}$ 
and $\mathcal{W}^\dagger$) defined in (\ref{QyW}),   
interchange the state vectors $\ket{n,k,k_3,\pm,\gamma}$ and 
$\ket{n,k,k_3,\pm,-\gamma}$ according to the rules in 
(\ref{qdonpsi}), (\ref{Qonphi+}) and (\ref{W+onpsi1}),
(\ref{W+onphi2}).
The
ground  
states of $\mathcal{H}$ ($\breve{\mathcal{H}}$) which 
are given by $\ket{n,k,k_3,-,-1}$ ($\ket{n,k,k_3,+,+1}$ ) 
are invariant under transformations generated 
by these fermionic operators, therefore the
quantum system $\mathcal{H}$ 
exhibits the unbroken
$\mathcal{N}=2$ Poincar\'e
supersymmetry. 
  
Finally, the spectrum generating ladder operators for the
supersymmetric system correspond to  
operators $\mathcal{G}$ and $\mathcal{G}^\dagger$  for $n$,
$\mathcal{K}_\pm$ for $k_3$ and 
the matrix non-local operators  
\begin{eqnarray}
\left(
\begin{array}{cc}
\mathcal{A}_\pm &0\\
0 & \tilde{\mathcal{A}}_\pm
\end{array}\right)\,,\qquad
\left(
\begin{array}{cc}
\mathcal{A}_\pm^\dagger &0\\
0 &  \tilde{\mathcal{A}}_\pm^\dagger
\end{array}\right)\,.\qquad
\end{eqnarray}
for the angular quantum number $k$.

\subsection{Dimensional reduction}
In this section we trace out how two different 
super-extensions 
of the one-dimensional AFF model 
can be obtained by a  reduction
 from our three-dimensional 
superconformal system.  
For the sake of simplicity we put here $\omega=1$, and denote $\sqrt{\omega}r=r$ as $x$.

Let us revisit first the supersymmetric AFF model. 
There are two possible extensions, which 
are given by the $2\cross2$ matrix Hamiltonians 
\begin{equation}
\label{superAFF}
\mathcal{H}_\ell^{\epsilon}=\begin{pmatrix}
H_\ell+\epsilon(\ell-\frac{1}{2}) & 0 \\
0 & H_{\ell-1}+\epsilon(\ell+\frac{1}{2})
\end{pmatrix}\,,\quad
H_\ell=\frac{1}{2}\left(-\frac{d^2}{dx^2}+x^2+\frac{\ell(\ell+1)}{x^2}\right)\,,
\end{equation}
where $\epsilon=\pm$ and $\ell\geq-1/2$ \cite{LLQM,FPW,FalPis,KirLoy}. 
The $\Z_2$-grading operator is $\sigma_3$, and 
the supercharges of  super-extended  systems $\mathcal{H}_\ell^{\epsilon}$ are given by 
\begin{equation}
Q^\epsilon_{\ell,1}=-\frac{\epsilon}{\sqrt{2}}\left(\begin{array}{cc}
0 & A^\epsilon_\ell\\
{A^\epsilon_\ell}^\dagger & 0
\end{array}\right)\,,\qquad
Q^\epsilon_{ \ell,2}=i\sigma_3 Q^\epsilon_{ \ell,1}\,,
\end{equation}
where 
\be
A^\epsilon_\ell=-\epsilon\frac{d}{dx}+x+\epsilon\frac{ \ell}{x}\,.
\ee
The supercharges and Hamilltonian operators  satisfy the $\mathcal{N}=2$ Poincar\'e 
superalgebra
\begin{equation}
\{Q^\epsilon_{\ell,a},Q^\epsilon_{\ell,b}\}=2\delta_{ab}\mathcal{H}^{\epsilon}_\ell\,,\qquad
[\mathcal{H}^{\epsilon}_\ell, Q^\epsilon_{\ell,a}]=0\,,\qquad a,b=1,2\,.
\end{equation} 
As in the case studied in the previous section, here we can also construct 
the 
$R$ symmetry generator 
\be\label{Rell}
R_\ell=\tfrac{1}{2}(\mathcal{H}^{-}_\ell-\mathcal{H}^+_\ell)=\tfrac{1}{2}\sigma_3-\ell\,,
\ee
and therefore one Hamiltonian can be 
expressed in terms of 
another 
and $R_\ell$.
Additionally, we have the conformal symmetry ladder operator 
\begin{equation}
\mathfrak{G}_\ell=\left(\begin{array}{cc}
G_{\ell} & 0\\
0 & G_{\ell-1}
\end{array}\right)\,,\qquad
G_{\ell}=-\frac{1}{2}\left(\frac{d}{dx}+x\right)^2+\frac{\ell(\ell+1)}{2x^2}\,,
\end{equation}
and its 
adjoint,
which are generated 
by 
\be\label{2rootG}
\{Q^\epsilon_{\ell,a},Q^{-\epsilon}_{\ell,b}\}=
\delta_{ab}(\mathfrak{G}_{\ell}+\mathfrak{G}_{\ell}^\dagger)
+
i\epsilon_{ab}(\mathfrak{G}_{\ell}-\mathfrak{G}_{\ell}^\dagger)\,.
\ee
 By constructing nilpotent fermionic operators 
$\frac{1}{2}(Q^{\epsilon}_{\ell,a}\pm iQ^{\epsilon}_{\ell,a})\,,$ 
it is not difficult to show that 
these generators satisfy the algebra (\ref{osp(2,2),1})-(\ref{osp(2,2),7}). 

The eigenstates of the super-Hamiltonian $\mathcal{H}^\epsilon_\ell$ 
and supercharge $Q^\epsilon_{\ell,1}$, 
 which we will denote as $\Phi_{n,\ell,\varrho}^{\epsilon}$ with 
  $\varrho=\pm1$,
are given by
\begin{eqnarray}
\label{satatesepsilon-1}
&
\Phi_{0,\ell,1}^{-}= \frac{x}{\sqrt{2}}\left(\begin{array}{c}
0\\
f_{0,\ell-1}(x)
\end{array}\right)\,,
&\\&
\Phi_{n+1,\ell,1}^{-}= \frac{x}{\sqrt{2}}\left(\begin{array}{c}
f_{n,\ell}(x)\\
f_{n+1,\ell-1}(x)
\end{array}\right)\,,\qquad
\Phi_{{n+1},\ell,-1}^{-}=\sigma_3
\Phi_{{n+1},\ell,1}^{-}
\label{phinj}\,, \label{satatesepsilon-2}
&\\&
\label{SpectralSuSY2}
\Phi_{n,\ell,1}^{+}= \frac{x}{\sqrt{2}}\left(\begin{array}{c}
f_{n,\ell}(x)\\
f_{n,\ell-1}(x)
\end{array}\right)\,,\qquad
\Phi_{n,\ell,-1}^{+}=\sigma_3\Phi_{n,\ell,1}^{+}\,,\qquad n\in\N_0\,.
\label{phinj2}
\end{eqnarray}  
The spectral equations are 
\begin{equation}
\label{SpectralSuSY1}
\mathcal{H}^\epsilon_{{\ell}}\Phi_{n,{\ell},\varrho}^{\epsilon}=(2n+\beta_{\pm}(2{\ell}+1))
\Phi_{n,{\ell},\varrho}^{\epsilon}\,,\qquad\
Q^\epsilon_{{\ell},1}\Phi_{n,{\ell},\varrho}^{-}=\varrho\sqrt{(2n+\beta_{\epsilon}(2{\ell}+1))}
\Phi_{n,{\ell},\varrho}^{\epsilon}\,,
\end{equation}
where $\beta_+=1$, $\beta_-=0$, cf.  (\ref{qdonpsi}).
  In the case of 
$\mathcal{H}_\ell^-$, the ground state $\Phi_{0,\ell,1}^{-}$
is annihilated by the super-Hamiltonian and by the supercharges $Q_{\ell,a}^-$,
and, therefore, 
supersymmetry is unbroken,
with energy levels being independent  of $\ell$.
On the other hand,  
 $\mathcal{H}_\ell^+$ has no
 zero-energy ground state,  energy levels depend on parameter $\ell$,
 and  there is no
physical eigenstate 
annihilated by both supercharges $Q_{\ell,a}^{+}$,
that implies that supersymmetry is spontaneously broken. 
For more details  see refs.
\cite{CIP,LM2,InzPly4}. 
The independence and dependence of energy levels on $\ell$
is reminiscent of two subsets of states 
in our three-dimensional  system with infinitely degenerate 
energy levels due to their independence on the quantum
number $j$ and finitely degenerate, depending on $j$ energy
eigenvalues. This is an additional indication
on that  one-dimensional superconformal extensions 
of the AFF model  (\ref{superAFF}) may indeed 
be obtained by reduction from our
three-dimensional   $\mathfrak{osp}(2\vert 2)$ superconformal
system.

In the following, we will show that for 
two different dimensional reductions of the system
$\mathcal{H}$ defined  in (\ref{superH}),  
we obtain a particular realization of the 
one-dimensional super-extended AFF model in  both,
broken and unbroken, $\mathcal{N}=2$ 
supersymmetry phases,
with $\ell=j$ taking one of the values 
$j=|\nu|,|\nu|+1\ldots$. To this end we first  note
that the Hamiltonian $\mathcal{H}$ admits a
 representation 
\begin{equation}
\label{HintermsRK}
\mathcal{H}=\frac{1}{2}\left[-\frac{1}{x^2}\frac{\partial}{\partial x}\left(x^2\frac{\partial}{\partial x}\right)
+ x^2\right]\mathbb{I}_{4\cross 4}+ \frac{1}{2x^2}(\mbfgr{\mathcal{K}}^2-\Gamma\mathcal{R}+\tfrac{3}{4})+\mathcal{R}\,.
\end{equation} 
Also, let us 
introduce
the following notation to distinguish one-dimensional from three-dimensional generators:  
\begin{eqnarray}
&
\mathscr{B}_{j,{\alpha}}^{\epsilon}=\{\mathcal{H}_j^\epsilon,{{R}_j},
\mathfrak{G}_j,\mathfrak{G}_j^\dagger\}\,,
\qquad 
\mathscr{F}_{j,{\beta}}^{\epsilon}=\{Q_{j,1}^{\epsilon},Q_{j,2}^{\epsilon},Q_{j,1}^{-\epsilon},Q_{j,2}^{-\epsilon}\}\,,
&\\&
\mathcal{B}_{{\alpha}}=\{\mathcal{H},{\mathcal{R}_j},\mathcal{G}_j,\mathcal{G}_j^\dagger\}\,,\qquad
\mathcal{F}_{{\beta}}=\{Re(\mathcal{Q}),Im(\mathcal{Q}),
Re(\mathcal{W}),Im(\mathcal{W})\}\,,&
\end{eqnarray}
where we imply that $\mathcal{B}_{j,1}=\mathcal{H}$ etc.,
and $Re(\mathcal{Q})=\tfrac{1}{2}(\cQ+\cQ^\dagger)$,
$Im(\mathcal{Q})=\tfrac{i}{2}(\cQ^\dagger-\cQ)$.

For the dimensional reduction we ``extract"  a subspace  in which the angular and spin operators 
in (\ref{HintermsRK}) take fixed numerical values. We have two independent choices which
we distinguish  by the signs $\pm$, and they relate to the
choice of the states 
$\ket{\chi,\pm}$ defined by the set of equations
\begin{eqnarray}
\label{reduction1}
&
\mbfgr{(\mathcal{K}}^2-k(k+1))\ket{\chi,\pm}=0\,,\qquad 
(\mathcal{K}_3-k_3)\ket{\chi,\pm}=0\,,\qquad 
&\\&
\label{reduction2}
\mathcal{P}_{\pm}\ket{\chi,\pm}=0\,,\qquad
\mathcal{P}_{\pm}=\frac{1}{2k+1}(\Pi_\pm+k\mp \mathcal{R})\,,
&
\end{eqnarray}
where 
$k=j\pm\tfrac{1}{2}\,,$ and  
$k_3=j_3\pm\tfrac{1}{2}\,$. Here, 
the most general form of $\ket{\chi,\pm}$ is 
\begin{equation}\label{anbn}
\ket{\chi,\pm}=\sum_{n=0}^{\infty}a_n^\pm\ket{n,k,k_3,\pm,1}+b_n^\pm \ket{n,k,k_3,\pm,-1}=
\sum_{n=0}^{\infty}
\left(\begin{array}{cc}
a_n^\pm \ket{n,k,k_3,\pm}\\
b_n^\pm \Vert n,k,k_3,\pm\rangle
\end{array}\right)\,.
\end{equation}
The operators $\mathcal{P}_{\pm}$ are projectors onto 
the orthogonal subspaces $\ket{\chi,-}$ and $\ket{\chi,+}$. In both subspaces, 
the grading operator preserves its form, 
while the action of  operators   
$\mathcal{R}$ and 
$\mathcal{K}^2-\Gamma\mathcal{R}+3/4$ produce 
\begin{eqnarray}
&
\mathcal{R}\ket{\chi,-}=\left(\begin{array}{cc}
-(j-\frac{1}{2})\mathbb{I}_{2\cross 2}&0 \\
0& -(j+\frac{1}{2})\mathbb{I}_{2\cross 2}
\end{array}\right)\,\ket{\chi,-}=R_{j}\otimes\mathbb{I}_{2\cross2}\ket{\chi,-}, &\\&
\mathcal{R}\ket{\chi,+}=\left(\begin{array}{cc}
(j+\frac{3}{2})\mathbb{I}_{2\cross 2}&0 \\
0& (j+\frac{1}{2})\mathbb{I}_{2\cross 2}
\end{array}\right)\,\ket{\chi,+}=-\sigma_1(R_{j+1})\sigma_1\otimes\mathbb{I}_{2\cross2}\ket{\chi,+},&\\&
(\mbfgr{\mathcal{K}}^2-\Gamma\mathcal{R}+\frac{3}{4})\ket{\chi,\pm}=\left(\begin{array}{cc}
j(j+1) & 0\\
0 &j(j\pm1)
\end{array}\right)\otimes\mathbb{I}_{2\cross 2}\,\ket{\chi,\pm},
\end{eqnarray} 
where the generator  $R_{j}$ defined in (\ref{Rell}) appears explicitly. In the same way 
we found in the subspace represented by $\ket{\chi,-}$ the following relations, 
\begin{eqnarray}
&
\mathcal{B}_{a}\ket{\chi,-}=\frac{1}{x}\mathscr{B}_{j}^{-}x\otimes\mathbb{I}_{2\cross 2}\ket{\chi,-}\,,\qquad
\mathcal{F}_{b}\ket{\chi,-}=\frac{1}{x}\mathscr{F}_{j}^{-}x\otimes\sigma_r\ket{\chi,-}\,,
&
\end{eqnarray}
while  in the subspace given by $\ket{\chi,+}$ we obtain
\begin{eqnarray}
&
\mathcal{B}_{a}\ket{\chi,+}=\sigma_1(\frac{1}{x}\mathscr{B}_{j+1}^{+}x)\sigma_1\otimes\mathbb{I}_{2\cross 2}\ket{\chi,+},
\quad
\mathcal{F}_{b}\ket{\chi,+}=\sigma_1(\frac{1}{x}\mathscr{F}_{j}^{+}x)\sigma_1\otimes\sigma_r\ket{\chi,+}.\quad
&
\end{eqnarray}
In these equations the generators take the form of a direct product 
of two  operators 
 $A\otimes B$, where 
  $A$ is a one-dimensional  $2\cross2$  matrix operator,  
and  $B$ is the  $2\cross2$
identity matrix or $\sigma_r$. 
The latter still contains an 
angular dependence, see (\ref{sigman}). To  
eliminate  the
 angular variables we introduce the operators 
\begin{equation}
\mathcal{O}_{\pm}=\left(\begin{array}{cc}
\ket{v}\bra{k,k_3,\pm} & 0\\
0 & \ket{v}\bra{k,k_3,\pm}\sigma_r\ 
\end{array}\right)\,,\qquad
\ket{v}=\left(\begin{array}{c}
1\\
1
\end{array}\right),
\end{equation}
and their adjoints. 
Here $\ket{k,k_3,\pm}$ 
corresponds to (\ref{Wspin+-}). Acting on the state $\ket{\chi,\pm}$,
 these operators produce
\begin{eqnarray}
\mathcal{O}_\pm\ket{\chi,\pm}=\ket{\Psi,\pm}\,,\qquad\bra{r}\ket{\Psi,\pm}=\sum_{n=0}^{\infty}
\left(\begin{array}{cc}
a_n^\pm f_{n,j}\\
a_n^\pm f_{n,j}\\
b_n^\pm f_{n,j\pm1}\\
b_n^\pm f_{n,j\pm1}\\
\end{array}\right)\,,\qquad
\end{eqnarray}
and $(\mathcal{O}_{\pm})^\dagger \ket{\Psi,\pm}=\ket{\chi,\pm}$,
that 
implies that  $(\mathcal{O}_{\pm})^\dagger\mathcal{O}_{\pm}\ket{\chi,\pm}=\ket{\chi,\pm}$ and
$\mathcal{O}_\pm(\mathcal{O}_\pm)^\dagger\ket{\Psi,\pm}=\ket{\Psi,\pm}$. 
Multiplication of the bosonic generators 
by $\mathcal{O}_\pm$ from the left 
and by $\mathcal{O}_\pm^\dagger$ from the right
does not change their
structure, i.e. 
$\mathcal{O}_{\pm}\mathcal{B}_{\alpha}(\mathcal{O}_{\pm})^\dagger=\mathcal{B}_{\alpha}$,
 but the same operation applied to fermionic generators 
produces 
\begin{eqnarray}
&
\mathcal{O}_{-}\mathcal{F}_{b}\mathcal{O}_{-}^\dagger\ket{\Psi,-}=
\frac{1}{x}\mathscr{F}_{j,\beta}^{-}x\otimes\sigma_1\,\ket{\Psi,-}\,,
&\\&
\mathcal{O}_{+}\mathcal{F}_{b}\mathcal{O}_{+}^\dagger\ket{\Psi,+}=
\sigma_1(\frac{1}{x}\mathscr{F}_{j+1,\beta}^{+}x)\sigma_1\otimes\sigma_1\,\ket{\Psi,+}\,.
&
\end{eqnarray}
Note that $\sigma_r$ disappears, 
and we effectively  eliminated 
the angular degrees of freedom. 
The reduction scheme is almost done. To
complete it  we introduce the unitary matrix 
 \begin{equation}
U=\left(\begin{array}{cccc}
1 & 0 & 0 &0 \\
0 & 0 & 1 & 0\\
0 & 1&0  &0 \\
0 & 0 & 0 & 1
\end{array}\right)\,,\qquad
UU^\dagger=1\,,\qquad
\text{det}\, U=-1\,,
\end{equation}  
which finally  gives  
 \begin{eqnarray}
 \label{finalreduction1}
&
U\mathcal{B}_a\ket{\Psi,-}=\frac{1}{x}\left(\begin{array}{cc}
\mathscr{B}_{j,a}^- & 0\\
0 & \mathscr{B}_{j,a}^-
\end{array}\right)x U\ket{\Psi,-}\,,&\\&
U\mathcal{O}_-\hat{\mathcal{F}}_\beta\mathcal{O}_{-}^\dagger \ket{\Psi,-}=
\frac{1}{x}\left(\begin{array}{cc}
0 &\mathscr{F}_{j,\beta}^- \\
\mathscr{F}_{j,\beta}^- & 0
\end{array}\right)x\,U\ket{\Psi,-}\,,
&\\&
U\mathcal{B}_a\ket{\Psi,+}=\frac{1}{x}\left(\begin{array}{cc}
\sigma_1\mathscr{B}_{j+1,a}^+\sigma_1 & 0\\
0 & \sigma_1\mathscr{B}_{j+1,a}^+\sigma_1
\end{array}\right)x U\ket{\Psi,+} \,,
&\\&
U\mathcal{O}_+\hat{\mathcal{F}}_\beta\mathcal{O}_{+}^\dagger \ket{\Psi,+}=
\frac{1}{x}\left(\begin{array}{cc}
0 &\sigma_1\mathscr{F}_{j+1,\beta}^+ \sigma_1\\
\sigma_1\mathscr{F}_{j+1,\beta}^+\sigma_1 & 0
\end{array}\right)x\,U\ket{\Psi,+}\,,
\label{finalreduction22}
\end{eqnarray} 
where each of these matrices  is a $4\cross 4$ matrix,  and 
\begin{eqnarray}
&
\bra{r}U\ket{\Psi,\pm}=\sum_{n=0}^{\infty}
\left(\begin{array}{c}
a_n^\pm f_{n,j}\\
b_n^\pm f_{n,j\pm1}\\
a_n^\pm f_{n,j}\\
b_n^\pm f_{n,j\pm1}\\
\end{array}\right)
\,.&
\end{eqnarray} 
The last state
contains two copies of the same two-component column vector,
which in turn can 
be expanded in terms of  eigenstates (\ref{satatesepsilon-1}), (\ref{satatesepsilon-2}) divided 
by $x$ in the case when  we do the reduction with sign $-$, or in terms of the states 
(\ref{phinj2}) multiplied by $\sigma_1/x$, if we choose the  sign $+$. On the other hand, 
in equations (\ref{finalreduction1})-(\ref{finalreduction22}) particular 
bosonic (fermionic) generators appear as block-(anti)diagonal matrices,
where
 each block corresponds to the same  one-dimensional generator.  To eliminate one of these copies 
 we can use the projector operators $\Pi_\pm$. Then we obtain 
 \begin{eqnarray}
 &
\bra{\vr}\Pi_\pm U\mathcal{B}_{\alpha}\ket{\Psi,-}\rightarrow\mathscr{B}_{j,\alpha}^{-}\Psi_{j}^-(x)\,,\qquad
\bra{\vr}\Pi_\pm U\mathcal{B}_{\alpha}\ket{\Psi,+}\rightarrow\sigma_1\mathscr{B}_{j+1,\alpha}^{+}\Psi_{j+1}^+(x)\,,
&
\quad
 \end{eqnarray}
 \begin{eqnarray}
& 
\bra{\vr}\Pi_\pm U\mathcal{O}_-\mathcal{F}_{\alpha}(\mathcal{O}_{-})^\dagger\ket{\Psi,-}\rightarrow
\mathscr{F}_{j,\beta}^-\Psi_{j}^-(x)\,, &\\&
\quad
\bra{\vr}\Pi_\pm U\mathcal{O}_+\mathcal{F}_{\alpha}(\mathcal{O}_{+})^\dagger\ket{\Psi,+}\rightarrow
\sigma_{1}\mathscr{F}_{j+1,\beta}^+\Psi^+_{j+1}(x)\,,\qquad &
 \end{eqnarray}
where
 \begin{eqnarray}
&
\Psi_{j}^{-}=x\bra{\vr}\Pi_{\pm}U\mathcal{O}_-\ket{\chi,-}=\sum_{n=0}^{\infty}A_{n}^-
\Phi_{n,j,1}^-+B_{n}^-\Phi_{n,j,-1}^-\,,
&\\&
\Psi_{j+1}^{+}=x\sigma_1\bra{\vr}\Pi_{\pm}U\mathcal{O}_{+}\ket{\chi,+}=\sum_{n=0}^{\infty}A_{n}^+
\Phi_{n,j+1,1}^++B_{n}^+\Phi_{n,j+1,-1}^+\,.
&
\end{eqnarray}
and the coefficients  $A_n^\pm$ and $B_n^\pm$
can be expressed in terms of $a_{n}^\pm$ and $b_n^\pm$ 
in (\ref{anbn}) 
using the orthogonality of states (\ref{satatesepsilon-1})-(\ref{SpectralSuSY2})
with $\ell=j$. 
 
Thus, the appropriately realized dimensional reduction of
our three-dimensional $\mathfrak{osp}(2\vert  2)$ superconformal system 
$\mathcal{H}$ with the  unbroken $\mathcal{N}=2$ Poincar\'e 
supersymmetry produces 
two different  $\mathfrak{osp}(2\vert  2)$
superconformal extensions 
of the one-dimensional AFF model with
unbroken or  spontaneously 
broken $\mathcal{N}=2$  Poincar\'e supersymmetries.

%%%%%%%%%%%%%%%%%%%%%%%%%%%%%%%%%%%
%%%%%%%%%%%%%%%%%%%%%%%%%%%%%%%%%%%%%%%%%%%%%

\section{Discussion and outlook}\label{DiscussionSection}
In summary, this work is  
divided  in two parts. 
In the first part, we studied the special case of a dynamical
 conformal system
presented by 
a  scalar charged particle 
in the monopole background  which is characterized 
by the presence of an additional,  
hidden symmetry that controls and reflects 
its peculiar classical and quantum properties.
In the second part, 
we added spin degrees of freedom by 
 introducing a spin-orbit coupling  of a special, unique 
 form that  guarantees a very peculiar degeneracy of energy 
 levels and gives rise to the superconformal
 $\mathfrak{osp}(2\vert  2)$ symmetry. By two different dimensional reduction schemes
 this three-dimensional supersymmetric system produces 
 the one-dimensional superconformal  extensions of 
 the AFF model \cite{AFF} in
 unbroken  and spontaneously 
 broken phases of $\mathcal{N}=2$ Poincar\'e 
 supersymmetry \cite{LM2}.

 The  scalar charged particle in the monopole background  that we
 considered 
 is subjected to a 
central potential 
$V(r)=\frac{\alpha}{2mr^{2}}+\frac{m\omega^2}{2}r^{2}$, which is  a 
 three-dimensional analog 
of the AFF model's potential, and
 therefore the system posseses the conformal Newton-Hooke symmetry 
 \cite{NH1,NH2,NH3,NH4}.
For coupling constant $\alpha\not=\nu^2$, 
trajectories are  closed only  for some particular 
 initial conditions. On the contrary,
 the special case $\alpha=\nu^2$ we study 
always gives us closed trajectories, 
the angular period is twice the radial period,
and even more, the dynamics projected 
to the plane orthogonal
to the Poincar\'e angular momentum vector $\vJ$ turns out
to be similar to  that for the usual three-dimensional isotropic 
harmonic oscillator. 
In fact, such 
an  interesting ``coincidence" is a
universal  property of the monopole 
background~\footnote{For earlier discussion of
the  quantum mechanical and classical aspects of such a universality 
see \cite{Zwan,Mardoyan,Mardoyan+}. We thank A. Nersessian
for drawing our attention to these works.
}.
Indeed, if we  consider the system described by the Hamiltonian 
\begin{equation}
\label{centralpotentialH}
H_\nu=\frac{\vpi^2}{2m}+\frac{\nu^{2}}{2mr^{2}}+ U(r)\,
\end{equation}
with  an arbitrary central potential $U(r)$,
then the dynamics of the vector variables 
$\vr\cross\vJ$ and $\vpi\cross\vJ$ has the same form  as that 
for vector variables 
 $\vr\cross\vL$ and $\vp\cross\vL$ when $\nu=eg=0$
 with $\vL$ being the usual angular momentum:  
\begin{eqnarray}
\begin{array}{|c|c|}
\hline
\nu\not=0 &\nu=0 \\\hline
\frac{d}{dt}(\vr\cross\vJ)=\frac{1}{m}\vpi\cross\vJ & \frac{d}{dt}(\vr\cross\vL) =\frac{1}{m}\vp\cross\vL\\\hline
\frac{d}{dt}(\vpi\cross\vJ)=U'(r)\,\vn\cross \vJ & 
\frac{d}{dt}(\vp\cross\vL)=U'(r)\,\vn\cross \vL\\
\hline
\end{array}\nonumber
\end{eqnarray} 
As a consequence, the motion in the plane  orthogonal to $\vJ$ is equivalent to the 
dynamics obtained in the absence of the monopole,  and if we 
know the solutions   $\vr=\vr(t)$ and $\vp=\vp(t)$ in the case $\nu=0$,
the dynamics 
for $\vpi\cross\vJ$ and $\vr \cross \vJ$ is at hand. To reconstruct the complete dynamics   
we combine the relations (\ref{relationJyr}) to obtain 
\begin{equation}
\vr(t)=\frac{1}{J^2}\left(\vJ\cross(\vr(t)\cross\vJ)+\sqrt{\frac{|\vr(t)\cross\vJ|}{J^2-\nu^2}}\vJ\right)\,.
\end{equation} 
In particular, if instead of (\ref{centralpotentialH}) we have a system
described by the Hamiltonian $\widetilde{H}_\nu=\frac{1}{2m}\vpi^2 +\widetilde{U}(r)$
with arbitrary central potential $\widetilde{U}(r)$, 
it is reduced in an obvious way to the system (\ref{centralpotentialH}) with central potential
$U(r)=\widetilde{U}(r)-\nu^2/2m r^2$.
The indicated similarity and relation allows, particularly, 
to identify immediately the analog of the  
Laplace-Runge-Lenz vector (\ref{LRL-G}) 
for a particle in the monopole background 
in the case of $\widetilde{U}=0$ and $U=0$,   and for the Kepler 
problem with $U=q/r$,  that was done 
earlier in \cite{Mono2,Mono2+,N4monopole} and 
\cite{Vinet} but by using a 
different approach.

From this perspective, one can  speculate that this 
peculiar dynamics should be related with the 
motion  of a particle  in a conical geometry under  the 
action of a potential $U(r)$,
or from the perspective  of gravity, 
with the dynamics 
 in a global monopole space-time \cite{GRmonopole}.
 In fact, generalizations of $SU(2)$ 
 systems with $D(2,1;\alpha)$ superconformal mechanics
 in Einstein-Maxwell background were studied recently  
 in \cite{GalLech}. It would be very interesting to
 generalize the system with harmonic trap 
that we considered for the case of $D(2,1;\alpha)$
superconformal mechanics and to look 
for its relation with the systems from \cite{GalLech}
in the light of the conformal bridge transformation.
In another but somehow related direction,  
it could be interesting to study this system and its 
hidden symmetries from the perspective 
of Eisenhart-Duval lift \cite{Cariglia2} and  Killing-Yano tensors
 \cite{Cariglia}.

The similarities in the dynamics are 
revealed not only at the classical level,
 but also in the quantum theory. In particular, the Hamiltonian operator 
 (\ref{Qm Hamiltonian}) 
 has the form of a three-dimensional harmonic oscillator Hamiltonian with a modified 
 angular momentum, which takes values $j=|\nu|+k$ with $k=0,1,\ldots$. Also, 
 degeneracy of the spectrum, related with the ratio of the classical radial and angular periods,
 can be explained in terms of the hidden integrals of motion 
 as we did in section \ref{algebraic approach}.
  Earlier results on the quantum analogy was obtained for this system 
 and for the Kepler potential  
 in \cite{Vinet} in the case of integer values of 
 $\nu$. 
 
 On the other hand,
 though the systems of the form (\ref{centralpotentialH}) with $U=0$ and $U=\tfrac{1}{2}m\omega^2r^2$ 
 classically and quantum mechanically are essentially different 
 since their Hamiltonians are generators 
 of conformal $\mathfrak{sl}(2,\R)$ symmetry of  non-compact
  and compact topological nature,  respectively, 
 they correspond to  
   two  different forms of dynamics
  governed  by conformal symmetry in the sense of Dirac \cite{Dirac,InzPly4}. 
 This fact allowed us to relate them at the quantum level 
 (that also can be done classically) by applying the conformal 
 bridge transformation \cite{Conformalbridge}, as we did this 
 in section \ref{Conformal bridge}.
 Symmetry generators
 of one of these systems, including those of hidden symmetry, 
 are mapped into symmetries of  the other system. 
  This transformation also allowed us to obtain the coherent states for 
  the system we studied.

In the second part,  similarly to the construction of 
the Dirac oscillator
\cite{DiracOs},  
we introduce additional spin degrees of freedom 
at the quantum level and adding a spin-orbit coupling term $\pm \omega' \vJ\cdot\vsigma$.
The constant value
$\omega'=\omega$ is very special as then 
the spectrum is divided in two subsets. The eigenvalues
in one subset do not depend on the quantum angular momentum  
number $j$ and hence are infinitely degenerate.
In the other subset each energy level has finite degeneracy
defined  by the constant  $\nu=eg$
which can only take integer and half-integer values.
Using the hidden symmetries of the scalar system, as well as its 
conformal Newton-Hooke  symmetry, we construct independent pairs of 
non-local ladder operators acting within both subspaces, 
one with infinite and one with finite degeneracy of energy 
levels. Here we do not compute commutators of these  
objects and the question on  the
symmetry algebra of the system remains unanswered.

The system with spin degrees of freedom gives rise to an $\mathcal{N}=2$ 
supersymmetric system characterized by the  
$\mathfrak{osp}(2\vert  2)$ superconformal symmetry.
Applying two different dimensional reduction schemes to the obtained  
superconformal system produces in one case the one-dimensional 
superconformal  extension of the AFF model with harmonic trap 
in the phase of the unbroken $\mathcal{N}=2$  Poincar\'e supersymmetry, 
while  in the other case gives us the same system 
but in the   spontaneously broken phase \cite{InzPly,CIP,LM2,InzPly4}. 
In this context, it would be interesting
to look for three-dimensional generalizations  of the 
one-dimensional rationally deformed superconformal 
systems constructed recently in \cite{CIP,LM2,InzPly4} by using 
dual Darboux transformations.

Hermitian supercharges of three-dimensional supersymmetric 
quantum mechanics can be related with ($3+1$)-dimensional 
Dirac operators in Euclidean space 
by setting  $\partial_t\rightarrow 0$,
and adding a gauge field connection. It is known 
that for self-dual or anti-self-dual electromagnetic fields
an extended $\mathcal{N}=4$ supersymmetry
can be obtained  \cite{Andreas1}.  
In the present case,   
the Hermitian combination $\mathcal{Q}^++\mathcal{Q}^-$, where $\mathcal{Q}^\pm$ are given in 
(\ref{QyW}), 
can be re-written 
in terms of Euclidean Dirac matrices
\begin{equation}
\gamma^{i}=\left(\begin{array}{cc}
0 & -i\sigma_i\\
i\sigma_i & 0
\end{array}\right)\,,\qquad
\gamma^{0}=\left(\begin{array}{cc}
0 & 1\\
1 & 0
\end{array}\right)\, 
\end{equation}
in the following form: 
\begin{equation}
\label{PB-Dirac Op}
\mathcal{Q}_{0}=-\sqrt{2}(\mathcal{Q}^++\mathcal{Q}^-)=\gamma^{i}(p_i-e\mathscr{A}_i)+e\gamma^{0}\mathscr{A}_0\,,
\end{equation}
where 
\begin{equation}
\mathscr{A}_{0}=\frac{g}{r}\,,\qquad
\mathscr{A}_{i}=A_{i}-i\frac{\omega}{e}\gamma^{5}\,r_i\,,
\end{equation}
with
 $\gamma^{5}=\Gamma$ is  our grading operator in section \ref{osp22 extension}. 
Then the operator (\ref{PB-Dirac Op}) can be viewed
as a parity breaking Euclidean Dirac operator with
components of the gauge potential satisfying the relations 
$-\partial_i \mathscr{A}_{0}=\epsilon_{ijk}\partial_{j}\mathscr{A}_{k}
=gr_i/r^3$.
Hence we are dealing with a new type of parity breaking dyon
background. Actually, the $\gamma^{5}$ terms do not allow for 
an
$\mathcal{N}=4$ supersymmetric extension and we
only have  $\mathcal{N}=2$ supersymmetry, with the 
second supercharge given by $i\sqrt{2}(\mathcal{Q}^+-\mathcal{Q}^-)=i\gamma^5\mathcal{Q}_{0}$.
It is 
interesting  
to relate a parity-breaking Dirac operator with a 
supersymmetric quantum mechanics.  
In this context it  is not clear whether  
a (pseudo)classical supersymmetric system exists 
whose quantization would produce our three-dimensional 
$\mathfrak{sl}(2,\R)$ 
superconformal system,  
or we have here a kind of 
a classical anomaly \cite{clasanomaly}.

To further interpret  the three-dimensional supersymmetric 
system, one can study limiting  cases of the coupling constant. 
In particular, in the limit $\nu\rightarrow 0$
we recover the non-relativistic limit of the Dirac oscillator  
considered in \cite{Balen,DiracOs,RevLett,QueMosh,Quesne}, and 
the supersymmetry (\ref{osp(2,2),1})-(\ref{osp(2,2),7}) remains
intact. On the other hand, in the limit, $\omega\rightarrow 0$, 
both Hamiltonians in (\ref{superH}) take the form 
\begin{equation}
\label{DionH}
\mathcal{H}_{\text{dyon}}=\frac{1}{2}
\left(\vpi^2+\frac{\nu^2}{r^2}-\frac{4\nu}{r^3}\,
\mathbfcal{S}_i^-\cdot\vr\right)\,,\qquad
\mathbfcal{S}^-=\frac{1}{2}(1-\Gamma)\mathbfcal{S}\,,
\end{equation} 
where $\mathbfcal{S}=\mathbb{I}\otimes\frac{1}{2}\vsigma$ 
denotes the vector  
spin  operator. This is just a Pauli type 
Hamiltonian for a charged spin-1/2 particle in a field of a self-dual 
dyon
\cite{N4monopole}. In the same limiting case
the operator (\ref{PB-Dirac Op}) is a Dirac type Hamiltonian 
which is identified as a supercharge related to (\ref{DionH}). 
As we have 
emphasized earlier, this system has extended
$\mathcal{N}=4$ supersymmetry. 
However, taking the limit $\omega\rightarrow0$ in our system  
(and following the approach in \cite{InzPly})
we cannot reconstruct the other three supercharges 
and one may suspect that something is still missing in
our construction. 
One possible way to answer this question is to try to perform a supersymmetric 
extension of the conformal bridge \cite{Conformalbridge}  and to apply this to 
the system  (\ref{DionH}). 

Finally, another 
interesting question related
to the Killing-Yano tensor problem mentioned above
is the possible existence of an additional,  hidden non-linear
supersymmetry 
in the system studied by us in section  \ref{osp22 extension}.
Such a possibility is suggested by the presence of such 
symmetries
in the system of a spin-1/2 particle in a self-dual dyon background \cite{N4monopole},
in superconformal mechanics at special values
of the boson-fermion coupling constant \cite{LeiPlyAna,LeiPlyAna+},
in the system of a scalar  particle
investigated by us in section \ref{hidclassym}, 
and the nonlinear exotic supersymmetry 
seen in the systems of spinning particles
in backgrounds  characterized by the presence of 
 Killing-Yano tensors \cite{GibRHol,Tanim,Plymonosusy,CarigliaYano,FroKrtKub}. 

 \vskip0.1cm
\noindent {\large{\bf Acknowledgements} } 
\vskip0.1cm

The work was partially supported by 
the CONICYT scholarship 21170053 (LI),
FONDECYT Project 1190842 and DICYT, USACH (MSP),
and the Project USA 1899 (MSP and LI).
LI and MSP also thank  FSU for  hospitality
during various stages of this  work.

\appendix

\section*{Appendix}

\section{The monopole harmonics}
\label{armonicosdemonopolo}
The monopole vector gauge  potential  possesses a singularity often called  Dirac string,  
because of  which  one may split the domain of definition of this field in two parts (charts) related by a gauge transformation.
The continuity conditions for the wave function 
in the transition region imply the remarkable result that $\nu=eg$ can take 
only integer and  half integer values at the quantum level. 
Here we
obtain an explicit expression 
for
 monopole harmonics, and 
 for this purpose it  is enough to work in 
 the fixed gauge   
\begin{equation}
\label{Gauge}
\vA=\frac{gz}{r(x^2+y^2)}(y\hat{x}-x\hat{y})=-\frac{g}{r}\cot(\theta)\hat{\varphi}\,,\qquad
\hat{\varphi}=(-\sin \varphi, \cos \varphi)\,.
\end{equation} 
To see how 
the monopole harmonics change under the corresponding 
gauge transformation, see ref. \cite{monoharm,monoharm+}. 
With this choice, the spherical components of the 
Poincar\'e vector $\vJ$ are given by 
\begin{equation}
J_{\pm}=e^{\pm i\varphi}\left(i\cot\theta\frac{\partial}{\partial\varphi}\pm
\frac{\partial}{\partial\theta}-\frac{\nu}{\sin\theta}
\right)\,,\qquad
J_{3} 
=-i\frac{\partial}{\partial \varphi}\,.
\end{equation}
These operators can be obtained as a ``reduction" of the  
$\mathfrak{so}(4)=\mathfrak{so}(3)\oplus\mathfrak{so}(3)$ symmetry
\begin{eqnarray}
&
[\mathscr{J}_i,\mathscr{J}_j]=i\epsilon_{ijk}\mathscr{J}_k\,,\qquad
[\mathscr{K}_a,\mathscr{K}_b]=i\epsilon_{abc}\mathscr{K}_c\,,\qquad
[\mathscr{J}_i,\mathscr{K}_a]=0\,,
\end{eqnarray}
 of the spinning top. 
Consider the following realization of this algebra 
\cite{Angularmomentum},
\begin{eqnarray}
&
\mathscr{J}_{\pm}=\mathscr{J}_{1}\pm i\mathscr{J}_2=e^{\pm i\varphi}\left(i\cot\theta\frac{\partial}{\partial\varphi}\pm
\frac{\partial}{\partial\theta}-\frac{i}{\sin\theta}\frac{\partial}{\partial\psi}
\right)\,,\qquad
\mathscr{J}_3=-i\frac{\partial}{\partial \varphi}\,,&\\&
\mathscr{K}_{\pm}=\mathscr{K}_{1}\pm i\mathscr{K}_2=e^{\mp i\psi}\left(i\cot\theta\frac{\partial}{\partial\psi}\mp
\frac{\partial}{\partial\theta}-\frac{i}{\sin\theta}\frac{\partial}{\partial\varphi}
\right)\,,\qquad
\mathscr{K}_3=i\frac{\partial}{\partial \psi}\,.
\end{eqnarray}
Here,  $\theta$, $\varphi$ and $\psi$ are the Euler angles, 
$0\leq\varphi,\psi<2\pi$,  $0\leq\theta<\pi$,
and $\mathscr{J}_i\mathscr{J}_i=\mathscr{K}_a\mathscr{K}_a=\mathscr{J}^2$.
The common eigenstates of  $\mathscr{J}^2$, $\mathscr{J}_3$ and $\mathscr{K}_3$
satisfying  relations
\begin{eqnarray}
&
\mathscr{J}^2D_{j,m,m'}=j(j+1)D_{j,m,m'}\,,\quad
\mathscr{J}_\pm D_{j,m,m'}=\sqrt{(j\mp m)(j\pm m+1)}D_{j,m\pm 1,m'}\,,&\\&
\mathscr{K}_\pm D_{j,m,m'}=\sqrt{(j\mp m')(j\pm m'+1)}D_{j,m,m'\pm1}\,,&\\&
\mathscr{J}_3D_{j,m,m'}=mD_{j,m,m'}\,,\qquad
\mathscr{K}_3 D_{j,m,m'}=m'D_{j,m,m'}\,&
\end{eqnarray}
are given by the generalized spherical functions 
\begin{equation}
\label{generalspherical}
D_{j,m,m'}(\varphi,\theta,\psi)=e^{i(m\varphi+m'\psi)}P_{j,m,m'}(\cos\theta)\,,
\end{equation}
where
\begin{eqnarray}
\label{Plmm'}
&
P_{j,m,m'}(u)=\mathcal{N}_{j,m,m'}(1-u)^{-\frac{m-m'}{2}}(1+u)^{-\frac{m+m'}{2}}
\left(\frac{d}{du}\right)^{j-m}(1-u)^{j-m'}(1+u)^{j+m'}\,,\nonumber &\\& 
\mathcal{N}_{j,m,m'}=\frac{(-1)^{j-m'}}{2^j(j-m')!}\sqrt{\frac{(j-m)!(j+m')!}{(j+m)!(j-m')!}}\,,&\\&
\nonumber 
j=0,1/2,1,3/2,\ldots\,,\qquad m,m'=j,j-1,\ldots,-j\,.&
\end{eqnarray}
The necessary reduction is achieved by imposing the condition
\be\label{K3psi}
(\mathscr{K}_3-\nu)\Psi(\theta,\varphi,\psi)=0
\ee
on a wave function $\Psi(\theta,\varphi,\psi)$ being a linear combination of the states
$D_{j,m,m'}(\varphi,\theta,\psi)$. This equation has a nontrivial solution only
when a constant parameter $\nu$ takes some integer or half-integer value
that corresponds to the Dirac quantization condition for  $\nu=eg$.
Fixing integer or half-integer value for $\nu$, a general solution of (\ref{K3psi}) 
is a linear combination of the states  $D_{j,m,\nu}$, with $j=\vert \nu \vert, \vert \nu \vert+1,\ldots$,
$m=-j,-j+1,\ldots, j$,  
and therefore the monopole harmonics are given by 
\begin{equation}
\mathcal{Y}_{j}^{j_3}(\theta,\phi;\nu)=e^{-i\nu\psi}D_{j,j_3,\nu}(\theta,\phi,\psi)\,. 
\end{equation}

\section{The derivation of $A_{n,j,m}$ and  $B_{n,j,m}$ }
\label{derivAB}

To clarify the action of operators $\va^\pm$ it is convenient to introduce 
notation
\begin{equation}
\label{sphericalcomponent}
\eta_0=a_3\,,\qquad \eta_{\pm}=a_1\pm i a_2
\,.
\end{equation}
Then  
\begin{eqnarray}
&
\label{vector-commutation1}
[H,\eta_q]=-\omega\eta_q\,, \qquad
[J_3,\eta_0]=[J_\pm,\eta_{\pm}]=0\,,\qquad
[J_3,\eta_{\pm}]=\pm\eta_{\pm}\,,
\qquad
 q=0,\pm\,, &\\&
\label{vector-commutation2}
[J_{\pm},\eta_0]=\mp\eta_{\pm }
\,,\qquad
[J_{\pm},\eta_{\mp}]=\pm 2\eta_{0}\,.
&
\end{eqnarray}
From equations (\ref{vector-commutation1}) one concludes that 
the action of $\eta_q$ and $\eta_q^\dagger$  
has to be of the form 
\begin{eqnarray}
\label{eta on psi}
&
\eta_q\ket{n,j,m}=A_{n,j,m}^{(q)}\ket{n,j-1,m+q}+B_{n,j,m}^{(q)}\ket{n-1,j+1,m+q}\,,
&\\&
\eta_q^\dagger\ket{n,j,m}=A_{n,j+1,m-q}^{(-q)*}\ket{n,j+1,m-q}+B_{n+1,j-1,m-q}^{(-q)*}\ket{n+1,j-1,m-q}\,.
&
\end{eqnarray}
The first equation in (\ref{vector-commutation2}) means that 
$A_{n,j,m}^{(\pm)}$ and $B_{n,j,m}^{(\pm)}$ are related with 
 $A_{n,j,m}^{(0)}\equiv A_{n,j,m}$ and $B_{n,j,m}^{(\pm)}\equiv B_{n,j,m}$ by means 
of the algebraic expressions 
\begin{eqnarray}
\label{ABpm}
\small{
\begin{array}{ll}
A_{n,j,m}^{(\pm)}=&\mp\sqrt{(j\mp m-1)(j\pm m)}A_{n,j,m}\pm\sqrt{(j\mp m)(j\pm m+1)}A_{n,j,m\pm1}\,,\\
B_{n,j,m}^{(\pm)} =&\mp\sqrt{(j\mp m+1)(j\pm m+2)}B_{n,j,m}\pm\sqrt{(j\mp m)(j\pm m+1)}B_{n,j,m\pm 1}\,,\\
\end{array}}
\end{eqnarray}
and using 
 the second equation in 
(\ref{vector-commutation2}) we derive the recurrence relations 
\begin{eqnarray}
\small{
\begin{array}{rll}
\frac{2(2j+j^2-m^2)B_{n,j,m}}{\sqrt{(j+1)^2-m^2}}&=&\sqrt{(j-m+2)(j+m+1)}B_{n,j,m-1}
+\sqrt{(j+m+2)(j-m)}B_{n,j,m+1},\\
2(j^2-m^2-1)A_{n,j,m} &=& [j^2-m^2]^{-\frac{1}{2}}(\sqrt{j^2-(m-1)^2}A_{n,j,m-1}+ \sqrt{(j+1)^2-m^2}A_{n,j,m+1}),\\
\end{array}}
\end{eqnarray}
the solutions of which are
\begin{equation}
A_{n,j,m}=\sqrt{(j+m)(j-m)}a_{n,j}\,,\qquad
B_{n,j,m}=\sqrt{(j+m+1)(j-m+1)}b_{n,j}\,.
\end{equation}     
To determine coefficients
$a_{n,j}$ and $b_{n,j}$ we use 
the relations  
\begin{equation}
[\eta_0,\eta_{\pm1}]=\mp \cC J_{\pm}
\qquad[\eta_0,\eta_0^\dagger]=\omega\big(2\vJ^2-J_{3}^2+1-\nu^2\big)\,,\label{cometazeta2}
\end{equation}
which produce the equations 
\begin{eqnarray}
\label{eq}
\small{
\begin{array}{rll}
(2j+3)b_{n,j}a_{n-1,j+1}-(2j-1)b_{n,j-1,m}a_{n,j}&=&\omega\sqrt{2n(2n+2j+1)}\,,\\
\omega(2j(j+1)+1-\nu^2-m^2)&=&(2j+1)(a_{n,j+1}^2-b_{n,j}^2)+\\
&&+(m^2-j^2)(a_{n,j}^2
-b_{n+1,j-1}^2+b_{n,j}^2-a_{n,j+1}^2).
\end{array}}
\end{eqnarray}
As $b_{n,j}$ and $a_{n,j}$  do not depend on $m$, the last equation implies the
identity 
$a_{n,j}^2-b_{n+1,j-1}^2+b_{n,j}^2-a_{n,j+1}^2=-\omega$. On the other hand,  
from equation (\ref{eta on psi}) with $q=0$ we conclude that the constant $b_{0,j}$ 
should vanish, contrary to the constant $a_{0,j}\not=0$. 
Using this and the first equation in (\ref{eq}) we have 
\begin{equation}
a_{n,j}=\tilde{a}_{n,j}\sqrt{\frac{\omega(j+\nu)(j-\nu)}{(2j-1)}}\,,\qquad
\tilde{a}_{0,j+1}=1\,. 
\end{equation}
Inserting this result and the anzatz 
\begin{equation}
b_{n,j}=\tilde{b}_{j}\sqrt{\frac{\omega2n(j+\nu+1)(j-\nu+1)}{(2j+1)}}\,,
\end{equation}
into equations Eq. (\ref{eq})  we finally obtain the system of equations 
 \begin{equation}
 (2j+3)\tilde{b}_{j}\tilde{a}_{n-1,j+1}=\sqrt{2n+2j+1}\,,\qquad
 2n\tilde{b}_{j}^2-\tilde{a}_{n,j+1}^2=-1
 \end{equation}
which has the solutions $\tilde{b}_{j}^2=(2j+3)$ and $\tilde{a}_{n,j}^2=(\frac{2n+2j+1}{2j+1})$. 
Collecting our results we end up with (\ref{Anlm1}).

\section{Generalized Laguerre polynomials}
\label{Laguerrerelations}
When acting with the first order operators
$\Theta,\Xi$ and their adjoint on the eigenspinors 
$\ket{n,k,k_3,\pm}$ and $\Vert n,k,k_3,\pm\rangle$,
the following functional relations for the
generalized Laguerre polynomials are useful:
\begin{align}
&y\frac{d}{dy}L^\alpha_n(y)-yL^\alpha_n(y)+\alpha L_n^\alpha=
(n+1)L_{n+1}^{\alpha-1}\,,\nonumber \\
&\frac{d}{dy}L_n^\alpha(y)-L_n^\alpha(y)=-L_n^{\alpha+1}(y)\,,
\nonumber \\
&\frac{d}{dy}L^{\alpha}_{n}(y)=-L_{n-1}^{\alpha+1}(y)\,,
\nonumber \\
&y\frac{d}{dy}L_n^\alpha(y)+\alpha L_n^\alpha(y)=
(n+\alpha)L_n^{\alpha-1}(y)\,.\label{laguerre_rec}
\end{align}

\section{Commutators  $[\mathcal{R},\mathcal{Q}]$ and $[\mathcal{R},\mathcal{W}^\dagger]$ }
\label{Apcommutator}
To compute these commutators, we first observe that
$\vb$ and $\vb^\dagger$ are vector operators with respect to $\vJ$, that means 
\begin{eqnarray}
&
\vJ\cdot\vb=\vb\cdot\vJ\,,\qquad
\vJ\cross\vb+\vb\cross\vJ=2i\vb\,.
&
\end{eqnarray}
Next we use the equality $\vpi\cross\vpi=i\frac{\nu}{r^2}\vn$ which implies
\begin{eqnarray}
&
\vJ\cdot\vb=\frac{1}{\sqrt{2}r}(i\nu-\nu\vr\cdot\vpi+i\omega\nu r^2)\,,\quad
\vJ\cdot\vb^\dagger=\frac{1}{\sqrt{2}r}(i\nu-\nu\vr\cdot\vpi-i\omega\nu r^2)\,.
&
\end{eqnarray}
With these identities and  representations of $\Theta$ and $\Xi$ 
in (\ref{Qnu}) and (\ref{Wsigmaform}), one can easily compute the
commutators
\begin{eqnarray}
&
[\mathcal{R},\mathcal{Q}]=\left(\begin{array}{cc}
0& (\vJ\cdot\vsigma+\frac{3}{2})\Theta+\Theta(\vJ\cdot\vsigma+\frac{3}{2}+2\nu\sigma_r)\\
0&0
\end{array}\right)=\mathcal{Q}\,,
 &\\&
[\mathcal{R}_\nu,\mathcal{W}^\dagger]=\left(\begin{array}{cc}
0& (\vJ\cdot\vsigma+\frac{3}{2})\Xi+\Xi(\vJ\cdot\vsigma+\frac{3}{2}+2\nu\sigma_r)\\
0&0
\end{array}\right)=\mathcal{W}^\dagger\,. 
&
\end{eqnarray}

\end{document}